# Experimental Quantum State Tomography of Optical Fields and Ultrafast Statistical Sampling

M. G. Raymer
Department of Physics, University of Oregon
Eugene, OR 97403
raymer@uoregon.edu

M. Beck
Department of Physics, Whitman College
Walla Walla, WA 99362
beckmk@whitman.edu

**Abstract**

We review experimental work on the measurement of the quantum state of optical fields, and the relevant theoretical background. The basic technique of optical homodyne tomography is described with particular attention paid to the role played by balanced homodyne detection in this process. We discuss some of the original single-mode squeezed-state measurements as well as recent developments including: other field states, multimode measurements, array detection, and other new homodyne schemes. We also discuss applications of state measurement techniques to an area of scientific and technological importance–the ultrafast sampling of time-resolved photon statistics.

# 1. Introduction

How can the quantum state of a physical system be completely determined by measurements? Before answering, it is useful to define what is ordinarily meant by a quantum state. Following Leonhardt, we affirm that "knowing the state means knowing the maximally available statistical information about all physical quantities of a physical object" [1]. Typically by "maximally available statistical information" we mean probability distributions. Hence, knowing the state of a system means knowing the probability distributions corresponding to measurements of any possible observable pertaining to that system. For multiparticle (or multimode) systems this means knowing joint probability distributions corresponding to joint measurement of multiple particles (or modes). Quantum mechanics is a theory of information, and the state of a system is a convenient means of describing the statistical information about that system. Interestingly, it has been shown that any extension of quantum theory in which the state (density matrix) does not contain all statistical information requires that an isolated system could receive information via EPR (Einstein-Podolski-Rosen) correlations, an effect which is widely viewed as nonphysical [2].

Since knowing the state means knowing all the statistical information about a system, is the inverse true? If one knows all of the statistical information about a system, does one then know (or can one infer) the quantum state of that system? Clearly a real experiment cannot measure all possible statistical information. A more practical question is then: can one infer reasonably well the quantum state of a system by measuring statistical information corresponding to a finite number of observables?

The answer to this question is a resounding yes, but there are some caveats. By now it is well established that the state of an individual system cannot, even in principle, be measured [3, 4]. This is easily seen by the fact that a single measurement of some observable yields a single value, corresponding to a projection of the original state onto an eigenstate that has nonzero probability. Clearly this does not reveal much information about the original state. This same measurement simultaneously disturbs the individual system being measured, so that it is no longer in the same state after the measurement. This means that subsequent measurements of this same system are no longer helpful in determining the original state. Since state measurement requires statistical information, multiple measurements are needed, each of which disturbs the system being measured. And attempting to use "weaker" measurements also does not help in this regard [5-7].

However, the state of an ensemble of identically prepared systems can be measured [3, 4]. Here each member of an ensemble of systems is prepared by the same state-preparation procedure. Each member is measured only once, and then discarded (or possibly recycled). Thus, multiple measurements can be performed on systems all in the same state, without worrying about the measurement apparatus disturbing the system. A mathematical transformation, of which there are several (see Refs. [1, 8-10] and the rest of this volume,) is then applied to the data in order to reconstruct, or infer, the state. The relevant interpretation of the measured state in this case is that it is the state of the ensemble. (Although we believe that any single system subsequently prepared by this same procedure will be described by this state as well.)

How many different observables need to be statistically characterized for an accurate reconstruction of the state? This is an interesting, and not easily answered, question. In general, the more complex the system, the more observables that are required. The minimum number is two; at least two noncommuting observables must be measured (see Royer [3] or Ballentine [4] and the references therein.) This is related to the "Pauli Question." Pauli asked in 1933 whether or not the wave function of a particle could be uniquely determined by measurements of its position and momentum distributions [11]. In three dimensions the answer turns out to be no. As pointed out by Gale et al., the probability distributions obtained from the position and momentum wave functions for the hydrogen atom $\psi(r,\theta,\varphi) = R_{n,l}(r) P_l^m(\theta) e^{im\varphi}$ are independent of the sign of $m$ [12]. However, note that the only difference between the wave functions with different signs of $m$ is that they are complex conjugates of each other. For pure states in one dimension, knowledge of the position and momentum distributions is sufficient to reconstruct the wave function up to a potential ambiguity in the sign of the complex phase function [13-15]. However, in the case of mixed states, even for one-dimensional systems, more than two distributions are needed for state reconstruction.

Because some of the first experimental work on quantum state measurement [16] analyzed the collected data using a mathematical technique that is very similar to the tomographic reconstruction technique used in medical imaging, and because all techniques are necessarily indirect, a generally accepted term for quantum state measurement has become quantum state tomography(QST.) Systems measured have included angular momentum states of electrons [17], the field [16, 18-20] and polarization [21] states of photon pairs, molecular vibrations [22], trapped ions [23], atomic beams [24], and nuclear spins [25]. Indeed, QST has become so prevalent in modern physics that it has been given its own Physics and Astronomy Classification Scheme (PACS) code by the American Institute of Physics: 03.65.Wj, State Reconstruction and Quantum Tomography.

The purpose of this article is to review some of the theoretical and experimental work on quantum state measurement. Prior reviews can be found in [1, 8, 9, 26]. We will concentrate here on measurements of the state of the field of an optical beam with an indeterminate number of photons in one or more modes. Measurement of the polarization state of beams with fixed photon number is discussed in the article by Altepeter, James and Kwiat in this volume. Nearly all field-state measurements are based on the technique of balanced homodyne detection, and hence fall under the category of optical homodyne tomography (OHT.) This technique was first suggested by Vogel and Risken [10] and first demonstrated by Smithey et al. [16]. A reasonably complete list of experiments that have measured the quantum states (or related quantities) of optical fields is presented in Table 1.

Here we will describe balanced homodyne detection, and its application to OHT. We emphasize pulsed, balanced homodyne detection at zero frequency (DC) in order to model experiments on "whole-pulse" detection. Such detection provides information on, for example, the total number of photons in a particular spatial-temporal mode [27]. (This differs from the other commonly used technique of radio frequency (RF) spectral analysis of the photocurrent [28-30].) Discussion of theoretical issues will be given in Secs. 2 and 3, while experimental issues will be discussed in Sec. 4.

Recently, array detectors have been used to perform QST [31-33]. One motivation for using arrays is to increase the effective detection efficiency, and hence the fidelity of the state reconstruction. Furthermore arrays allow measurements of the states of multiple modes, spatial or temporal, of the same beam. In Sec. 5 we review the use of array detection in QST.

| Ref. | Measured Quantities | # of modes | Description |
|---|---|---|---|
| 16 | $W$ | 1 | first implementation of OHT, squeezed vacuum |
| 34 | $W$, $\Pr(\phi)$ | 1 | quantum phase distributions |
| 35 | $W$, $\Psi$, $\langle x|\hat{\rho}|x'\rangle$, $\Pr(\phi)$ | 1 | first quantum wave function (vacuum) |
| 36 | *moments* | 1 | number-phase uncertainty relations |
| 37 | $W$, $\Psi$, $\Pr(n)$, $\Pr(\phi)$ | 1 | number-phase uncertainty relations, wave function of coherent state |
| 38 | *moments* | 1 | time resolved measurement |
| 39 | $\Pr(n)$ | 1 | quantum sampling, time resolved measurement |
| 40 | $W$ | 1 | first OHT with RF detection, squeezed vacuum |
| 41 | *moments* | 1 | time & frequency resolution |
| 42 | $\langle n|\hat{\rho}|m\rangle$ | 1 | even-odd oscillations of $P(n)$ |
| 43 | $g^{(2)}(t,t+\tau)$ | 2 | first two-mode measurement |
| 18 | $W$, $\langle n|\hat{\rho}|m\rangle$ | 1 | squeezed coherent state |
| 44 | $W$ | 1 | modulated signal |
| 45 | $\Pr(n)$ | 1 | self-homodyne tomography |
| 46 | $W$, $\Pr(n)$ | 1 | squeezed light from cold atoms |
| 19 | $W$ | 1 | first photon counting measurement |
| 47 | $\Pr(n,m)$ | 2 | joint photon number distributions |
| 48 | $\Pr(n)$ | 1 | application to telecommunications |
| 49 | W | 1 | self-homodyne tomography |
| 31 | $\langle n|\hat{\rho}|m\rangle$ | 2 | first array detection experiment |
| 20 | $W$, $\Pr(n)$ | 1 | one-photon state |
| 50 | *moments* | 2 | time-resolved polarization correlations |
| 51 | $W$, $\langle n|\hat{\rho}|m\rangle$ | 1 | high repetition rate |
| 33 | *Q, moments* | many | Q-function, unbalanced array detection |
| 52 | $W$, $\Pr(n)$ | 1 | very high repetition rate |
| 53 | $W$, $\langle n|\hat{\rho}|m\rangle$ | 1 | displaced one-photon state |
| 54 | $W$, $\Pr(n)$ | 1 | displaced one-photon state |
| 55 | $W$ | 1 | self-homodyne tomography, amplitude squeezing |
| 56 | $W$, $\Pr(n)$ | 1 | microwave one-photon state, Rydberg atoms |
| 57 | $\Pr(n,m)$ | 2 | universal homodyne tomography |
| 32 | $\langle n|\hat{\rho}|m\rangle$ | 2 | mode-optimized array detection |

Table 1 Experiments measuring quantum states and other related quantities.

**1.a Balanced Homodyne Detection and Optical Quantum State Tomography**

Since the development of balanced optical homodyne detection (BHD) by Yuen and Chan in 1983 [58, 59], it has been used widely in both continuous-wave [28, 29] and pulsed-laser [60] applications for characterizing the quadrature-amplitude fluctuations of weak optical fields. By quadrature amplitudes one means the cosine and sine, or real and imaginary, components of the oscillating electric field associated with a particular spatial-temporal mode. Schematically we can write the (real) field as

$$\begin{aligned} E &= E_{VAC}\left[q\cos(\omega t+\theta)+p\sin(\omega t+\theta)\right] \\ &= E_{VAC}\operatorname{Re}\left[(q+ip)\exp(-i\omega t-i\theta)\right] \end{aligned}, \quad (1.1)$$

where $E_{VAC}$ is an electric field corresponding to the vacuum (zero-point) strength of a single mode, and $q, p$ are the dimensionless quadrature amplitudes associated with the carrier frequency $\omega$ and reference phase $\theta$. The goal of BHD is to allow a quantum-limited measurement of these field amplitudes. This is a quite different strategy than the more common intensity or photon-flux detection that is accomplished by *directly* illuminating a photoemissive detector (such as a photodiode or a photomultiplier) with the light beam to be measured.

The idea of indirect quantum measurements is an outgrowth of the development of quantum-state reconstruction [39, 41, 61, 62]. Rather than targeting the quantum state itself as the object of measurement, one can target various statistical averages (moments) of the optical field, such as mean field, mean intensity, variance of field, or variance of intensity. When considering time-nonstationary or multimode fields, one can target various time-dependent correlation functions, which often are the quantities of greatest interest in quantum optics. Using BHD allows one to measure one type of quantity (field quadratures) and from their ensemble measurements infer the averages of other quantities. This offers a new way to view quantum measurements in general [1, 8].

There have been many theoretical discussions of balanced homodyne detection [63-77]. The measurement of the full probability distribution of the quadrature amplitudes was demonstrated in 1993 [16, 35], generalizing earlier experiments which measured only the variances of the quadrature amplitudes. The distributions were measured using pulsed, balanced homodyne detection with integrating (DC) photodetectors, a technique that allows the statistical characterization of weak, repetitive optical fields on ultrafast time scales [38, 39]. This technique detects optical field amplitudes in a particular spatial-temporal mode that is defined by a coherent local-oscillator (LO) pulse. Ideally the LO pulse should have a known spatial-temporal shape and should be phase-locked to the signal field, although useful information can be obtained even when there is no stable phase relation between LO and signal [39].

Combined with the data processing method of tomographically reconstructing the Wigner quasi-distribution for the two quadrature-field amplitudes, the method is called optical homodyne tomography [16, 35]. Alternatively one can reconstruct the density matrix directly from the measured quadrature distributions, as first pointed out by D'Ariano et al. [61] and Kuhn et al. [78].

The Wigner distribution $W(q,p)$ is related to the density matrix in the quadrature or number representations by [79]

$$W(q,p) = \frac{1}{2\pi} \int_{-\infty}^{\infty} \langle q + \frac{1}{2}q' | \hat{\rho} | q - \frac{1}{2}q' \rangle e^{-ipq'} dq'$$
$$= \sum_{n=0}^{\infty} \sum_{m=0}^{\infty} \langle n | \hat{\rho} | m \rangle W_{nm}(q,p)$$

. (1.2)

To obtain the form of the functions $W_{nm}(q,p)$, insert $\hat{\rho} = \sum_{n,m} \rho_{nm} |n\rangle\langle m|$ into the above, giving

$$W_{nm}(q,p) = \frac{1}{2\pi} \int_{-\infty}^{\infty} \psi_n(q + \frac{1}{2}q') \psi_m(q - \frac{1}{2}q') e^{-ipq'} dq' \quad , \tag{1.3}$$

where $\psi_n(q) = \langle q | n \rangle$ are the real wave functions of the number states. These are well known to be Hermite-Gaussian functions for a harmonic oscillator, such as an optical mode. The explicit forms for $W_{nm}(q,p)$ are Gaussian-Laguerre functions [1, pg. 129].

Clearly, knowing either the Wigner distribution or the density matrix in any representation is equivalent to knowing the quantum state of the ensemble of identically prepared systems (having a single degree of freedom) that have undergone measurement.

The Wigner distribution plays a natural role in the quantum theory of homodyne detection, since its marginal distributions give directly the measured quadrature-amplitude statistics [10] [see Eq. (3.3) below.] Because of this property, the Wigner distribution can be determined mathematically by tomographic inversion of the measured photoelectron distributions. In the case of a classical-like field (as from an ideal laser) the Wigner distribution can be interpreted simply as the joint distribution for the field quadrature amplitudes. In the case of a quantized field, it is uniquely related to the quantum state (density matrix or wave function) of a spatial-temporal mode of the signal field.

Mathematically, the quadrature probability distributions for different reference phases $\theta$ are given by the density matrix in the number (Fock) basis according to [1, p.118]

$$\Pr(q,\theta) = \sum_{\mu,\nu} \rho_{\mu\nu} \psi_\mu(q) \psi_\nu(q) \exp[i(\nu-\mu)\theta] \qquad . \qquad (1.4)$$

To understand this formula intuitively, note that the phase $\theta$ serves a role similar to a time variable in the Schrödinger time evolution of an energy eigenstate of a harmonic oscillator:

$$\psi_n(q,t) = \psi_n(q,0)\exp[-in\omega t] \;\leftrightarrow\; \psi_n(q,\theta) = \psi_n(q,0)\exp[-in\theta] \qquad . \qquad (1.5)$$

Then using $\langle q,\theta|n\rangle = \psi_n(q,\theta) = \psi_n(q)\exp(-in\theta)$ in $\Pr(q,\theta) = Tr[\hat{\rho}|q,-\theta\rangle\langle q,-\theta|]$ gives Eq. (1.4).

The idea behind quantum state tomography for a harmonic oscillator (or optical homodyne tomography for a single field mode) is that complete knowledge of $\Pr(q,\theta)$, through an ensemble of measurements on identically prepared systems, allows an accurate determination of the density matrix (or equivalently the Wigner distribution) – that is, the quantum state [1, 10, 16, 80]. This idea has also been extended to apply to systems other than harmonic oscillators [22, 81, 82].

The inversion of the raw measured data to arrive at a believable form of the quantum state is a delicate and interesting procedure. The inversion methods fall into two broad classes–deterministic and nondeterministic. In the deterministic methods an experimentally determined $\Pr(q,\theta)$ is used to determine, for example, $\rho_{nm}$, by a direct mathematical inversion of Eq. (1.4) [or Eq.(3.9) below.] But this approach begs the question, "How well can we determine or, more properly, infer the probability densities $\Pr(q,\theta)$ from a set of measurements of the quantities $q$ for many values of $\theta$?" This is essentially a problem of classical statistics, and the fact that its answer plays a critical role in quantum state estimation is intriguing. This would seem to provide a link between the micro-quantum world and the macro-classical world, as emphasized in early discussions, notably by Neils Bohr. For a recent discussion see Caves et al. [83].

The nondeterministic approach recognizes this question from the beginning and targets the estimation of the quantum state directly, rather than the classical distributions as intermediate objects. A broad perspective on this approach has recently been given by Schack, Brun, and Caves [84]. These authors (following K.R.W. Jones before them [85, 86]) consider a number $N+M$ of like systems, all prepared by the same procedure, one at a time. They describe the situation this way:

"Suppose one is given a (prior) state $\hat{\rho}^{(N+M)}$ on [Hilbert space] $H^{(N+M)}$ and the results of measurements on $M$ subsystems. The task is to find the (posterior) state of the remaining $N$ subsystems conditioned on the measurement results."[84]

Because all systems are prepared by the same (but unknown to us) procedure, the density operator of the entire ensemble prior to the measurements is entangled, and can be represented uniquely in the form

$$\hat{\rho}^{(N+M)} = \int d\hat{\rho}\, p(\hat{\rho})\hat{\rho}^{\otimes(N+M)} \qquad , \qquad (1.6)$$

where $p(\hat{\rho})$ is the "probability" that all systems are prepared in the (common) state $\hat{\rho}$, and the functional integral is over all possible prior density operators (which can be parameterized by a set of density-matrix elements). Consider a measurement on the first subsystem that yields a result "k," with probability $p_k$.

Schack et al. show that following such a measurement; the state of the remaining $N+M$-1 subsystems is conditioned to

$$\hat{\rho}^{(N+M-1)} = \int d\hat{\rho}\, p(\hat{\rho}|k)\, \hat{\rho}^{\otimes(N+M-1)} \quad , \quad (1.7)$$

where

$$p(\hat{\rho}|k) = \frac{p(k|\hat{\rho})}{p_k} \quad , \quad (1.8)$$

with $p(k|\hat{\rho})$ being the conditional probability to observe $k$ if the state is $\hat{\rho}$. Equation (1.8) has precisely the form of the Bayes Rule in classical probability or inference theory [87]. For this reason Schack et al. call this the Quantum Bayes Rule. The strategy in using this theorem for quantum state estimation is to iterate the procedure, updating the state after each measurement on a different system (which is then discarded), until only $N$ systems remain. To continue their description:

"If the measurements on individual subsystems correspond to an informationally complete POVM (positive-operator-valued measure) or if they contain sequences of measurements of a tomographically complete set of observables [88], the posterior probability on density operators approaches a delta function in the limit of many measurements. This is the case of quantum state tomography [16], which can thus be viewed as a special case of quantum Bayesian inference. In this limit, the exact form of the prior probability on density operators becomes irrelevant." [84]

This result provides a satisfying link between the deterministic and nondeterministic methods for QST. It leads to the conclusion that measurements of quadrature probability densities for many (ideally all) phase values is a well-founded scheme for accurately inferring the state of a large ensemble of identical and similarly prepared single-mode EM fields.

Nevertheless, there still remains an issue about the most effective strategy for state reconstruction when the number of measurements is not large enough for the limit suggested above to be valid. In particular, it has been observed that when the sample size is small, so that statistical uncertainties prevent obtaining highly accurate values for $\Pr(q,\theta)$, then the density matrix reconstructed using the deterministic inversion method can have nonphysical properties–for example, negative diagonal values (or probabilities) One satisfactory way to "cover up" these negative values is to calculate statistical error bars for each element of the density matrix [89]. Other methods rely on least-squares estimation [90], maximum-entropy estimation, [91], or maximum-likelihood estimation of the density matrix elements [92]. We will not present a review of all of these methods here. For reviews see [1, 8, 9, 93] as well as other chapters in this volume.

**1.b Why Measure the State?**

Aside from the question of how to measure the state of a quantum system, it is useful to ask why one would *want* to measure the state of a system. Of what use is knowing the state? An answer to this question is that once the state is obtained, distributions or moments of quantities can be calculated, even though they have not been directly measured. Indeed, one can calculate moments of quantities that do not correspond to Hermitian operators, and thus cannot even in principle be directly measured.

For example, some early measurements demonstrated the ability to measure photon number distributions at the single-photon level, even though the detectors used had noise levels that were too large to directly measure these distributions [37]. Distributions of optical phase have been measured, even though there is no known experimental apparatus capable of directly measuring this phase directly [34, 37]. Expectation values of the number-phase commutator $[\hat{n},\hat{\phi}]$ have been measured, even though this operator is not Hermitian, and thus cannot be directly observed [37].

Furthermore, even if the full quantum state is not measured, the same basic idea of performing many measurements corresponding to different observables can yield important information about optical fields.

For example, one can obtain information, with high time-resolution (on the time scale of 10's of fs), on the photon statistics of light propagation in scattering media [38], light emitted by edge-emitting lasers [39], or vertical-cavity surface-emitting lasers [50] and the performance optical communication systems [48, 94] We will discuss some of this work here.

## 2. Balanced Homodyne Detection of Temporal-Spatial Modes

Balanced homodyne detection, shown in Fig. 1, has the advantage that it rejects classical intensity fluctuations of the local oscillator (LO) field while measuring the signal quadrature-field amplitude in a particular spatial-temporal mode. Such a mode is defined by the space-time form of the LO pulse [64]. This provides a powerful technique for ultrafast time-gated detection (of field rather than intensity), recently developed into a practical scheme called *Linear Optical Sampling* [38, 41, 94]. The time resolution can be as short as the LO pulse (down to a few fs), while providing spectral resolution consistent with the time-frequency uncertainty product [41]. Further, because BHD measures the field quadrature amplitudes, for various phase values, it provides a tomographically complete set of variables (a "quorum," [95]) for quantum state tomography.

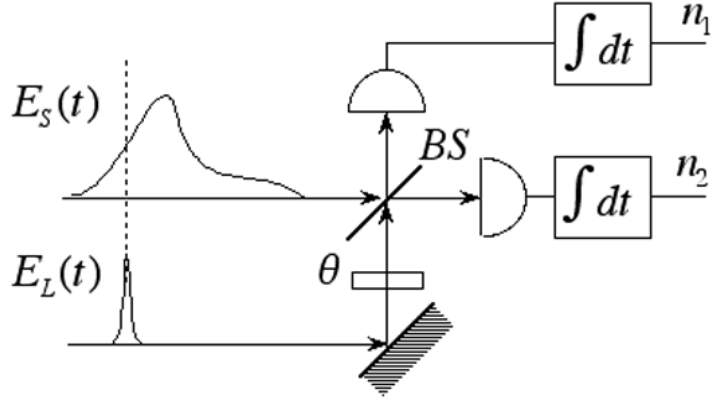

Fig. 1 BHD setup. A short local oscillator (LO) pulse $E_L$, after undergoing a phase shift theta, interferes with a signal pulse $E_S$ on a 50/50 beam splitter (or equivalent). The interfered pulses are detected by photodiodes and integrated to produce photoelectron numbers $n_1$ and $n_2$.

In homodyne detection the signal and LO fields have spectra centered at the same, or nearly the same, optical frequency. The signal electric-field $\hat{\underline{E}}_S^{(+)}$ interferes with the LO field $\hat{\underline{E}}_L^{(+)}$ at a loss-less beam splitter to produce two output fields, represented by vector-field operators [96-98],

$$\hat{\underline{E}}_1^{(+)} = t_1 \hat{\underline{E}}_S^{(+)} + r_2 \hat{\underline{E}}_L^{(+)}$$
$$\hat{\underline{E}}_2^{(+)} = r_1 \hat{\underline{E}}_S^{(+)} + t_2 \hat{\underline{E}}_L^{(+)} \quad , \qquad (2.1)$$

where unitarity of the transformation requires that the transmission and reflection coefficients satisfy the Stokes relation, $r_1 t_2^* + t_1 r_2^* = 0$. This guarantees that the output field operators commute. We assume the phase convention $t_1 = -t_2 = t$, $r_1 = r_2 = r$, and assume that $t$ and $r$ have values $(1/2)^{1/2}$, corresponding to a 50/50 beam splitter (effects of imperfect balance are discussed in [27].)

The LO field is usually assumed to be a strong, pulsed coherent field from a laser (nonclassical LO's can be treated [27].) The fields incident on the photodiode detectors generate photoelectrons with detection probability (quantum efficiency) equal to $\eta$, and the resulting current is integrated using low-noise charge-sensitive amplifiers [99, 100]. This provides the DC-detection, and differs from the often-used method of radio-frequency spectral analysis of the current to study noise at higher frequencies [59]. The values of the integrated photocurrents are recorded with analog-to-digital converters (ADC's), to give the numbers of photoelectrons $n_1$ and $n_2$ per pulse, which are then subtracted, pulse-by-pulse. Alternatively the photocurrents may be subtracted using an analog circuit and then converted to digital form [51, 99]. As seen below, the difference of these numbers is proportional to the chosen electric-field quadrature amplitude of the signal, spatially and temporally averaged over the space-time window defined by the duration of the LO pulse.

The positive-frequency part of the electric field operator is (underbars indicate vectors and carets indicate operators)

$$\hat{\underline{E}}_S^{(+)}(\underline{r},t) = i \sum_j \sqrt{\frac{\hbar \omega_j}{2\varepsilon_0}} \ \hat{b}_j \ \underline{u}_j(\underline{r}) \exp(-i\omega_j t) \quad . \tag{2.2}$$

Optical polarization is also indicated by the index $j$. In Dirac's quantization scheme the modes are plane waves, $\underline{u}_j(\underline{r}) = V^{-1/2} \underline{\varepsilon}^{(j)} \exp(i\underline{k}_j \cdot \underline{r})$, defined in some large volume $V$ (which may be taken to infinity later). Alternatively they can be taken as any complete set of monochromatic solutions to the source-free Maxwell equations, assumed to be orthogonal and normalized over volume $V$. The annihilation operators obey the commutator

$$[\hat{b}_j, \hat{b}_k^\dagger] = \delta_{jk} \quad . \tag{2.3}$$

We assume that the photodetectors respond to the incident photon flux. (This approach is different from the Glauber/Kelly-Kleiner formulation where the observed photoelectron current is in terms of the electromagnetic energy flux at the detector [101, 102].) The approach used here is considered to be more appropriate for broadband fields in the case that the detector's quantum efficiency is frequency independent, i.e., the detector is an ideal photoemissive, rather than an energy-flux, detector [63, 64, 66, 67, 70, 103, 104]. Appropriately, "photon-flux amplitudes" $\hat{\underline{\Phi}}_S^{(+)}(\underline{r},t)$ and $\hat{\underline{\Phi}}_L^{(+)}(\underline{r},t)$ can be defined for the signal and LO as

$$\hat{\underline{\Phi}}_S^{(+)}(\underline{r},t) = i\sqrt{c} \sum_j \hat{b}_j \ \underline{u}_j(\underline{r}) \exp(-i\omega_j t)$$

$$\hat{\underline{\Phi}}_L^{(+)}(\underline{r},t) = i\sqrt{c} \sum_j \hat{b}_{Lj} \ \underline{u}_j(\underline{r}) \exp(-i\omega_j t) \quad . \tag{2.4}$$

The photon fluxes (photon/sec) of the fields ($i$=1,2) at the detector faces at $z = 0$ are then represented by

$$\hat{I}_i(t) = \int_{Det} d^2x \ \hat{\underline{\Phi}}_i^{(-)}(\underline{x},0,t) \cdot \hat{\underline{\Phi}}_i^{(+)}(\underline{x},0,t) \quad , \tag{2.5}$$

integrated over the detectors' faces, where $\underline{x}$ denotes the transverse variables ($x, y$). It is assumed that the quantities $\hat{\underline{\Phi}}^{(+)}(\underline{r},t)$ obey the same transformations at the beam splitter as do the fields in Eq. (2.1). The photon number contained in a time interval (0,$T$) at each detector is represented by $\hat{N}_i = \int_0^T \hat{I}_i dt$. The interval duration $T$ is assumed to be longer than the duration of the signal and LO pulses, so that all of the energy of each is detected. The electronic system, including detector and amplifiers used to process the photocurrent, is assumed to act as a low-pass filter with an integration time (inverse bandwidth) much larger than $T$. The difference-photon number contained in a time interval (0,$T$) is represented by

$$\hat{N}_- = \hat{N}_1 - \hat{N}_2 = \int_0^T (\hat{I}_1 - \hat{I}_2) dt \quad , \tag{2.6}$$

and the total photon number is

$$\hat{N}_+ = \hat{N}_1 + \hat{N}_2 = \int_0^T (\hat{I}_1 + \hat{I}_2)\,dt \qquad . \qquad (2.7)$$

In the case of perfect balancing ($r = t$), the difference number is given by

$$\hat{N}_- = \int_0^T dt \int_{Det} d^2x\ \hat{\underline{\Phi}}_L^{(-)}(\underline{x},0,t)\cdot \hat{\underline{\Phi}}_S^{(+)}(\underline{x},0,t) \ + \ h.c. \qquad . \qquad (2.8)$$

As previously mentioned, this (DC) implementation of balanced homodyne detection is different than that in which a radio frequency (RF) spectrum analyzer is used to detect the oscillations of photocurrent at some frequency other than zero. The non-DC case can be analyzed by inserting into Eq.(2.6) and Eq.(2.7) a time-domain filter function centered at some nonzero frequency [67, 68, 71].

The quantum theory of BHD with pulsed LO's is closely related to the general theory of electromagnetic field quantization in terms of non-monochromatic, or wave packet, modes, first discussed by Titulaer and Glauber [105], and later used in the theory of Raman scattering [106]. Here we apply this approach to the photon-flux amplitude rather than the EM field. This choice brings a simplification of the formalism regarding the orthogonality of the wave packets. Rewrite the signal photon-flux amplitude as

$$\hat{\underline{\Phi}}_S^{(+)}(\underline{r},t) = i\sqrt{c}\sum_k \hat{a}_k\ \underline{v}_k(\underline{r},t) \qquad , \qquad (2.9)$$

where the wave packet modes are given by the superpositions

$$\underline{v}_k(\underline{r},t) = \sum_j C_{kj}\ \underline{u}_j(\underline{r})\exp(-i\omega_j t) \qquad , \qquad (2.10)$$

and $C_{kj}$ is a unitary matrix of coefficients. The new photon annihilation operators for the wave packet modes are related to the original operators by $\hat{a}_j = \sum_m C_{jm}^* \hat{b}_m$, and obey $[\hat{a}_j, \hat{a}_k^\dagger] = \delta_{jk}$. As an example of their use, a one-photon wave packet state is created by acting on the vacuum by $\hat{a}_k^\dagger |vac\rangle$.

The wave packet modes $\underline{v}_k(\underline{r},t)$ are orthonormal in the same large volume $V$ as are the original modes, $\int d^3r\ \underline{v}_k^*(\underline{r},t)\cdot \underline{v}_m(\underline{r},t) = \delta_{km}$. But because photodetection takes place in a plane (the surface of the detector) and not an infinite volume, we find it useful to reformulate the orthogonality of the wave packet modes. Consider that the detector plane (assumed infinite) is at $z = 0$, with transverse variables $\underline{x} = (x,y)$. Then it can be shown [see proof in Appendix] that for large time $T$ and paraxial (small-divergence) beams, the integral $c\int_0^T dt \int_{Det} d^2x$ acts similarly to the spatial volume integral $\int_0^{cT} dz \int_{Det} d^2x$, since the beam sweeps out such a volume in the detector integration time $T$. Therefore if we choose the $z$-dimension of the quantization volume to have length equal to $cT$, (with $T$ large), then we have orthogonality in the transverse-space-plus-time domain,

$$c\int_0^T dt \int_{Det} d^2x\ \underline{v}_k^*(\underline{x},0,t)\cdot \underline{v}_m(\underline{x},0,t) = \delta_{km} \qquad , \qquad (2.11)$$

where $\underline{v}_m(\underline{x},0,t)$ is the wave packet mode in the $z = 0$ plane. This property makes it easy to analyze BHD with a pulsed LO.

Let us assume that the LO pulse is a strong coherent state of a particular localized wave packet mode $\underline{v}_L(\underline{r},t)$. Then it is useful to separate the LO operator into a term for this mode and terms for all other (vacuum) modes,

$$\hat{\underline{\Phi}}_L^{(+)}(\underline{r},t) = i\sqrt{c}\ \hat{c}_L\ \underline{v}_L(\underline{r},t)\ +\ vacuum\ terms \qquad , \qquad (2.12)$$

where $\hat{c}_L$ is the annihilation operator for LO mode.

Inserting this into Eq.(2.8) and dropping the vacuum terms, which are small compared to the terms involving the strong LO, gives for the difference number

$$\hat{N}_- = -i\sqrt{c}\ \hat{c}_L^\dagger \int_0^T dt \int_{Det} d^2x\ \underline{v}^*_L(\underline{x},0,t)\cdot \hat{\underline{\Phi}}_S^{(+)}(\underline{x},0,t)\ +\ h.c.$$
$$= \hat{c}_L^\dagger \hat{a} + \hat{c}_L\hat{a}^\dagger \qquad , \qquad (2.13)$$

where

$$\hat{a} = \sum_k \hat{a}_k\ c \int_0^T dt \int_{Det} d^2x\ \underline{v}^*_L(\underline{x},0,t)\cdot \underline{v}_k(\underline{x},0,t) \qquad . \qquad (2.14)$$

This illustrates the spatial and temporal gating of the signal field, since it is multiplied by the LO field, which has some controlled shape–where the LO is zero, that portion of the signal is rejected.

If we assume that the detector is large enough to capture the whole transverse profile of the modes of interest, then the integral in Eq. (2.13) acts like an integral over the quantization volume, and orthogonality [Eq.(2.11)] shows that $\hat{a}$ in Eq.(2.14) stands for $\hat{a}_{k=L}$, the photon operator for the mode of the signal beam that has the same spatial-temporal shape as does the LO mode. It is the operator for the detected part of the signal field–the part that is "mode-matched" to the local oscillator. We can represent $\hat{a}$ by its "real" and "imaginary" parts given by $\hat{q} = (\hat{a} + \hat{a}^\dagger)/2^{1/2}$ and $\hat{p} = (\hat{a} - \hat{a}^\dagger)/i2^{1/2}$. These two variables are called quadrature-amplitudes for the detected spatial-temporal mode and obey $[\hat{q}, \hat{p}] = i$.

If the LO field is coherent and strong, and treated classically, with amplitude $\hat{c}_L$ in Eq.(2.13) replaced by $\alpha_L = |\alpha_L|e^{i\theta}$, then the difference number becomes [107]

$$\hat{N}_- = \hat{N}_-(\theta) = |\alpha_L|(\hat{a}e^{-i\theta} + \hat{a}^\dagger e^{i\theta}) \qquad . \qquad (2.15)$$

With the LO phase equal to zero, a balanced homodyne detector measures the real quadrature $\hat{q}$, while with the LO phase equal to $\pi/2$ it measures the imaginary quadrature $\hat{p}$. For arbitrary phase it measures the generalized quadrature amplitude

$$\hat{q}_\theta = \hat{N}_-(\theta)/(|\alpha_L|2^{1/2}) = (\hat{a}e^{-i\theta} + \hat{a}^\dagger e^{i\theta})/2^{1/2} \qquad . \qquad (2.16)$$

A conjugate variable is defined by

$$\hat{p}_\theta = (\hat{a}e^{-i\theta} - \hat{a}^\dagger e^{i\theta})/i2^{1/2} \qquad . \qquad (2.17)$$

These variables may be expressed as a rotation of the original variables,

$$\begin{pmatrix}\hat{q}_\theta\\ \hat{p}_\theta\end{pmatrix} = \begin{pmatrix}\cos\theta & \sin\theta\\ -\sin\theta & \cos\theta\end{pmatrix}\begin{pmatrix}\hat{q}\\ \hat{p}\end{pmatrix} \qquad . \qquad (2.18)$$

The variables $\hat{q}_\theta$ and $\hat{p}_\theta$ do not commute, and so cannot be measured jointly with arbitrarily high precision. Their standard deviations obey the uncertainty relation, $\Delta q_\theta \Delta p_\theta \geq 1/2$.

The photoelectron counting distribution for the difference photoelectron number $n_-$ in DC pulsed homodyne detector is [27]

$$\Pr(n_-) = \left\langle :\exp[-\eta(\hat{N}_1 + \hat{N}_2)]\left(\frac{\hat{N}_2}{\hat{N}_1}\right)^{n_-/2} I_{|n_-|}[(2\eta(\hat{N}_1\hat{N}_2)^{1/2}]:\right\rangle_{S,L} \qquad , \qquad (2.19)$$

where the double dots indicate normal operator ordering (annihilation operators to the right of creation operators), and $I_n(x)$ is the modified Bessel function of $n$-th order. This general result incorporates multimode, pulsed signal and LO fields in arbitrary quantum states. It can also accommodate non-mode-matched background from the signal beam (see [27] for a more general discussion.)

We consider the LO field to be in an intense coherent state $|\alpha_L\rangle$ of a single spatial-temporal-mode such that $\hat{c}_L |\alpha_L\rangle = \alpha_L |\alpha_L\rangle$. Then, the expression Eq.(2.19) can be evaluated and well approximated by a Gaussian function [27],

$$\Pr(n_-,\theta) = \left\langle : \frac{\exp[-(n_- - \eta \hat{N}_-(\theta))^2 / 2\eta \hat{N}_+]}{\left[\pi 2\eta \hat{N}_+\right]^{1/2}} : \right\rangle_S \quad , \quad (2.20)$$

where the difference number $\hat{N}_-(\theta)$ is given by Eq.(2.15) and the total number of photons hitting both detectors is

$$\hat{N}_+ = |\alpha_L|^2 + \hat{a}^\dagger \hat{a} \quad . \quad (2.21)$$

The total number of photons sets the scale for the shot-noise level. The quantum expectation value in Eq.(2.20) is with respect to the signal field, denoted by $S$, i.e., $\langle ... \rangle_S = Tr(\hat{\rho}_S ...)$ where $\hat{\rho}_S$ is the density operator for the signal mode(s).

We use the quadrature-amplitude operator $\hat{q}_\theta = \hat{N}_-(\theta)/|\alpha_L|^{2^{1/2}}$ defined in Eq.(2.16), and define the corresponding real variable $q_\theta = n_-/\eta |\alpha_L|^{2^{1/2}}$, (accounting for detector efficiency $\eta$). Then we can transform Eq.(2.20) into the probability density for quadrature amplitude of the mode-matched signal [27]

$$\Pr(q_\theta,\theta) = \left\langle : \frac{\exp[-(q_\theta - \hat{q}_\theta)^2 / 2\sigma^2]}{\left[\pi 2\sigma^2\right]^{1/2}} : \right\rangle_S \quad , \quad (2.22)$$

where $2\sigma^2 = 1/\eta$.

The distributions $\Pr(q_\theta,\theta)$ are experimentally estimated by repeatedly measuring values of $q_\theta$ for various fixed LO phases $\theta$ and building measured histograms $\Pr_M(q_\theta,\theta)$ of relative frequencies of the occurrence of each quadrature value. From these measured quadrature distributions one reconstructs the quantum state of the mode-matched signal field.

## 3. Quantum State Reconstruction and Optical Mode Statistical Sampling

### 3.a Inverse Radon Reconstructions

It can be shown that the (exact) quadrature distributions are related to the Wigner distribution of the signal mode by [27]

$$\Pr(q_\theta,\theta) = \int \int \frac{\exp[-\{q_\theta - \tilde{q}_\theta(q,p)\}^2 / 2\varepsilon^2]}{\sqrt{\pi 2\varepsilon^2}} W(q,p)\, dq\, dp \quad , \quad (3.1)$$

where $\tilde{q}_\theta(q,p) = q\cos\theta + p\sin\theta$ and $2\varepsilon^2 = 2\sigma^2 - 1 = 1/\eta - 1$. In the limit that $\eta = 1$, Eq.(3.1) becomes

$$\Pr(q_\theta,\theta) = \int \int \delta(q_\theta - \tilde{q}_\theta(q,p)) W(q,p)\, dq\, dp \quad . \quad (3.2)$$

Using $q = q_\theta \cos\theta - p_\theta \sin\theta$, and $p = x_\theta \sin\theta + p_\theta \cos\theta$, gives

$$\Pr(q_\theta,\theta) = \int_{-\infty}^{\infty} W(q_\theta \cos\theta - p_\theta \sin\theta, q_\theta \sin\theta + p_\theta \cos\theta)\, dp_\theta \quad . \quad (3.3)$$

This integral has the form of a marginal distribution, that is, the joint distribution has been integrated over one independent random variable to yield the distribution for the other variable.

The integral Eq.(3.3) is known as the Radon transform [1, 108], and has the form of a projection of the $W$ function onto the rotated $q_\theta$ axis, as illustrated by the line integrals in Fig. 2.

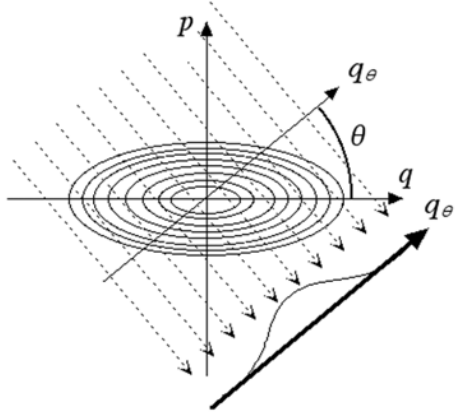

Fig. 2 Equal-value contours of a Gaussian-like Wigner distribution, showing squeezing. Line integrals perpendicular to an axis rotated by angle $\theta$ create a projected, or marginal, distribution $\Pr(q_\theta, \theta)$.

As pointed out in the context of quantum mechanics first by Bertrand and Bertrand [80] and independently by Vogel and Risken [10], Eq.(3.3) can be inverted to yield $W(q,p)$, given a set of distributions $\Pr(q_\theta, \theta)$ for *all* values of $\theta$ between 0 and $\pi$. The formal inversion is

$$W(q,p) = \frac{1}{4\pi^2} \int_{-\infty}^{\infty} dq_\theta \int_{-\infty}^{\infty} d\xi |\xi| \int_0^\pi d\theta \Pr(q_\theta,\theta) \exp[i\xi(q_\theta - q\cos\theta - p\sin\theta)] \quad . \quad (3.4)$$

In the case that one has measured a sufficient number of distributions $\Pr(q_\theta,\theta)$ for a finite set of discrete $\theta$ values, the inversion can be carried out numerically (with a certain amount of smoothing of the final result) using the well-studied filtered back projection transformation familiar in medical tomographic imaging [1, 108, 109]. This is the origin of the expression *optical homodyne tomography* [16], and later *quantum state tomography*. Equation (3.4) explicitly demonstrates the validity of the statements made in Sec. 1; namely that many measurements of many observables ($q_\theta$, for many values of $\theta$) enables one to determine the quantum mechanical state of a system [here the state is represented by $W(q,p)$.]

As mentioned above, if the inverse Radon transform Eq.(3.4) is applied to the measured histograms $\Pr_M(q_\theta,\theta)$, which are only estimates of the true distributions, nonphysical forms of the reconstructed state may erroneously appear. Error bars can be estimated in this case [89], or nondeterministic methods can be adopted [90, 92].

If the detector efficiency $\eta$ is less than unity, then the reconstruction does not yield the Wigner distribution of the signal mode [27, 78, 110]. It can be shown that the experimentally reconstructed Wigner distribution in this case is smoothed, and is given by

$$W_{Exp}(q,p) = \frac{1}{\pi 2\varepsilon^2} \int \int \exp[-(q-q')^2/2\varepsilon^2 - (p-p')^2/2\varepsilon^2] W(q',p') dq' dp' \quad , \quad (3.5)$$

where again $2\varepsilon^2 = 1/\eta - 1$. In principle, the smoothing function could be de-convolved from the measured distribution $W_{Exp}(q,p)$ to yield $W(q,p)$, but with experimental data having finite signal-to-noise ratio, and perhaps containing systematic errors, this is not usually practical. Although $W(q,p)$ can be negative, for $\eta \leq 0.5$ the integral in Eq.(3.5) is always positive definite, and so the measurement cannot show certain highly quantum effects.

If the signal field is excited in a *single* spatial-temporal mode $\underline{w}_S(\underline{r},t)$, different than that of the LO, $\underline{v}_L(\underline{r},t)$, then a part of it contributes to the detected amplitude, and the remainder to the background. In this case the detector efficiency $\eta$ is replaced by the product $\eta\, \eta_{LS}$, where the complex mode overlap is

$$\eta_{LS} = \left| c \int_0^T dt \int_{Det} d^2 x\ \underline{v}_L^*(\underline{x},0,t) \cdot \underline{w}_S(\underline{x},0,t) \right| \quad , \tag{3.6}$$

and $0 \leq \eta_{LS} \leq 1$. The mode-overlap factor $\eta_{LS}$, which includes both spatial and temporal overlap, acts like an additional attenuation factor for the mode of interest.

The first such state reconstruction was carried out by Smithey et al. in 1993, for the case of a squeezed state created by optical parametric amplification of the vacuum [16, 35]. Figure 3 shows the reconstructed density matrix in the quadrature representation, obtained by an inverse Fourier transform of the measured Wigner distribution [see Eq (1.2)]. The overall efficiency $\eta\, \eta_{LS}$ in this case was around 0.5, so the measured Wigner distribution Eq.(3.5) contained a fair amount of broadening and smoothing.

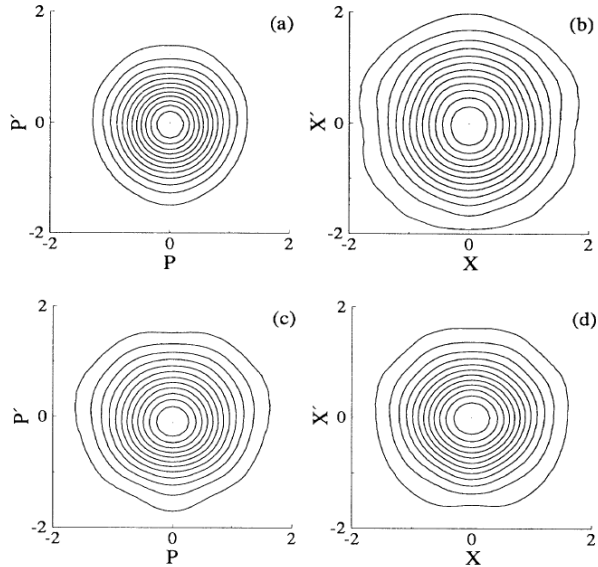

Fig. 3. Measured density matrix for (a), (b) the squeezed vacuum and (c), (d) the vacuum state, in q (called here x) or p representations: (a), (c) $\langle p+p'|\hat{\rho}|p-p'\rangle$, (b), (d) $\langle q+q'|\hat{\rho}|q-q'\rangle$. [16]

The first state reconstruction showing a negative Wigner distribution was performed by Lvovsky et al. (Fig. 4) [20]. Near-pure-state single-photon wave packets were created by parametric down-conversion combined with strong spatial and spectral filtering. An overall measurement efficiency of 0.55 allowed the one-photon component to dominate the vacuum component sufficiently to lead to a negative value of $W_{Exp}(0,0)$.

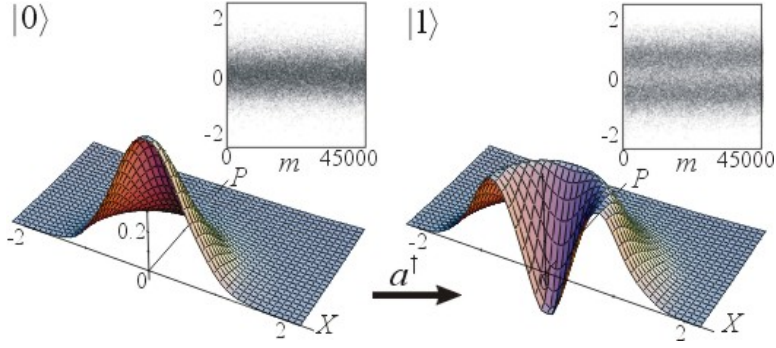

Fig. 4. Experimental reconstruction of a vacuum-state and a one-photon Wigner function. Insets show raw quadrature histogram data. [20 and private communication, A. Lvovksy]

If the state determined by OHT is determined to be a pure state (by computing $Tr[\hat{\rho}^2]$ and comparing it favorably to unity), then it is possible to reconstruct a Schrödinger wave function using

$$\psi(q) = \frac{\langle q | \hat{\rho} | q'=0 \rangle}{\sqrt{\langle q=0 | \hat{\rho} | q'=0 \rangle}} \, e^{i\beta_0} , \quad (3.7)$$

where $\beta_0$ is a nonphysical constant phase. An example for a coherent-state optical mode is shown in Fig. 5. (Note that because coherent states are robust against detector losses, it is valid to consider the measured quadratures as simply scaled down versions of the original quadratures. In this case the state that is measured is interpreted as that *after* suffering detector losses.)

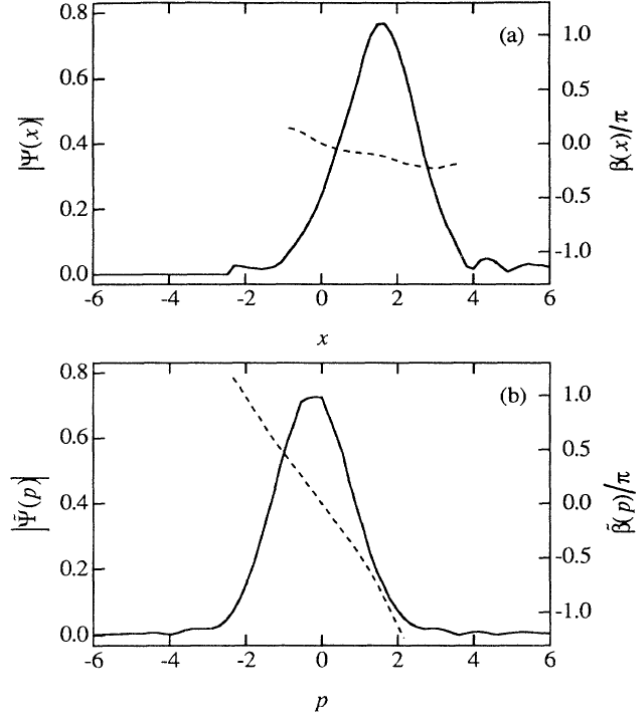

Fig. 5 Measured Schrödinger wave function of a coherent state of light with an average of 1.2 photons in (a) the q (or x)-quadrature representation and (b) the p-quadrature representation. [37] Solid line is the magnitude $|\psi|$ plotted, against the left axis; dashed lined is phase, plotted against the right axis.

Given an experimentally reconstructed Wigner distribution $W_{Exp}(q,p)$, one can invert Eq. (1.2) to obtain the measured density matrix in the quadrature $\langle q|\hat{\rho}|q'\rangle$ (as in Fig. 3) or number $\langle n|\hat{\rho}|m\rangle$ representations. From this density matrix one can compute probability distributions and statistical averages of any quantity of interest pertaining to the system studied.

A significant example of this type of indirect measurement is the reconstruction of the photon-number probability distribution $\Pr(n) = \langle n|\hat{\rho}|n\rangle$, which is obtained by measuring quadratures, not photon number directly. Also of interest is the distribution of the quantum phase $\phi$ of the signal field. Among the many useful definitions of quantum phase distributions [111], here we illustrate the idea using the London/ Pegg-Barnett distribution [112],

$$\Pr_{L/PB}(\phi) = \frac{1}{2\pi} \sum_{n,m=0}^{s} e^{i(m-n)\phi} \langle n|\hat{\rho}|m\rangle \quad , \quad (3.8)$$

where $s$ is a truncation parameter. We were the first to demonstrate the reconstruction of photon-number [37] and phase distributions [34, 35] for coherent and quadrature-squeezed states. We further demonstrated how to reconstruct the expectation value of the commutator of the number operator and the Pegg-Barnett phase operator, $[\hat{\phi}, \hat{n}]$ [36, 37]. This is interesting because this commutator is not a Hermitian operator and so does not correspond to a conventional "observable" in quantum theory; nevertheless its measurement (or rather the measurement of its moments) is amenable to the OHT technique.

Figure 6 shows the number-phase uncertainty product and the expectation value of the number-phase commutator, both determined from experiment for a coherent state with varying mean photon number. The uncertainty principle is of course satisfied. The interesting features are the degree to which the product does not equal the commutator, and the amount by which the product is less than 0.5, the limit that is expected in the limit of large photon numbers.

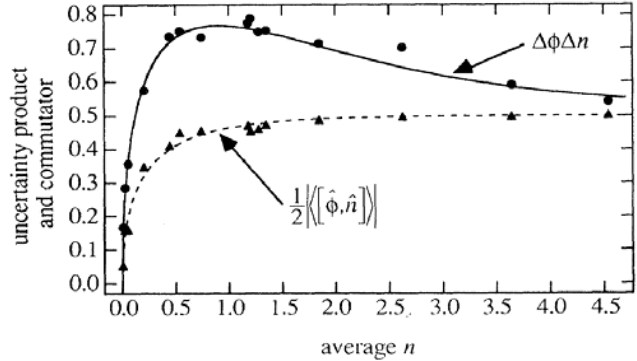

Fig. 6. Measured number-phase uncertainty product and expectation value of the number-phase commutator for a coherent state with mean number $n$. Continuous curves are theoretical predictions [37].

## 3.b Pattern Function Reconstructions

The path just described for obtaining the density matrix $\rho_{mn} = \langle m|\hat{\rho}|n\rangle$ and photon number distribution $\Pr(n)$ from the raw quadrature data is rather tortuous, and can introduce unnecessary error, in part as a result of the smoothing that is necessary in the inverse Radon transform. It was first pointed out by D'Ariano et al. that it is possible, and perhaps preferable, to bypass the inverse Radon transform [61]. Why not invert Eq. (1.4) directly to obtain the number-basis density matrix directly for the homodyne data? It was independently proposed that the quadrature-basis density matrix could be similarly obtained [78]. Although these methods do not use the Radon inverse, they have nevertheless come to be called quantum state "tomography" (from the original *optical homodyne tomography* [16]), as now are all schemes that use raw statistical data of one form or another to infer a quantum state.

The idea behind inverting Eq. (1.4) is rather simple when viewed from the perspective of dual bases. Rewrite Eq.(1.4) as

$$\Pr(q,\theta) = \sum_{\mu,\nu} \rho_{\mu\nu} G_{\mu\nu}(q,\theta) \quad , \quad (3.9)$$

where $G_{\mu\nu}(q,\theta) = \psi_\mu(q)\psi_\nu(q)\exp[i(\nu-\mu)\theta]$ serves as a basis for expanding the function $\Pr(q,\theta)$. The basis $\{G_{\mu\nu}(q,\theta)\}$ is comprised of linearly independent but nonorthogonal functions. Their linear independence guarantees that there exists a dual basis $\{F_{mn}(q,\theta)\}$, also nonorthogonal, with the bi-orthogonality property

$$\int_0^{2\pi} \frac{d\theta}{2\pi} \int_{-\infty}^{\infty} dq\, F_{mn}^*(q,\theta) G_{\mu\nu}(q,\theta) = \delta_{m\mu}\, \delta_{n\nu} \quad . \quad (3.10)$$

Using this, the relation Eq.(3.9) can easily be inverted to yield

$$\rho_{mn} = \int_0^{2\pi} \frac{d\theta}{2\pi} \int_{-\infty}^{\infty} dq\, F_{mn}^*(q,\theta) \Pr(q,\theta) \quad . \quad (3.11)$$

To find the form of the dual-basis functions, make the ansatz $F_{mn}(q,\theta) = M_{mn}(q)\exp[-i(m-n)\theta]$, and define $\chi_{\mu\nu}(q) = \psi_\mu(q)\psi_\nu(q)$, which, when inserted into Eq.(3.10), leads to the sole requirement [113]

$$\int_{-\infty}^{\infty} M_{n+D,n}(q)\, \chi_{\nu+D,\nu}(q)\, dq = \delta_{n\nu} \quad , \quad (3.12)$$

where $D = \mu - \nu = m - n$ is the common difference between the indices. This shows that the problem breaks up into a set of independent problems–one for each value of $D$. The dual functions can be found for each value of $D$ by defining the matrix of overlap integrals, $\wp_{\mu\nu}^{(D)} = \int_{-\infty}^{\infty} \varphi_\mu^{(D)}(q)\, \chi_{\nu+D,\nu}(q)\, dq$, where $\varphi_\nu^{(D)}(q)$ is a set of bases (one for each D), which can be nonorthogonal and may be chosen for convenience. It is easy to see that the dual-basis vectors are constructed as

$$M_{\mu+D,\mu}(q) = \sum_{\nu=\Omega}^{\infty} \left[\wp^{(D)-1}\right]_{\mu\nu} \varphi_\nu^{(D)}(q) \quad , \quad (3.13)$$

where $\wp^{(D)-1}$ is the inverse of the matrix $\wp^{(D)}$ and $\Omega$ is the greater of 0 and $-D$. A convenient choice for the numerical basis is given by a product of the Hermite-Gaussian functions and a Gaussian function, $\varphi_\nu^{(D)}(q) = (2\nu+1)^L \psi_{m(\nu)}(q)\exp(-q^2/2)$, where $m(\nu)$ is constrained to be even (odd) if $D$ is even (odd). $L$ is a parameter of order unity that can be adjusted to enhance numerical stability. This form allows the overlap integrals to be carried out in closed form.

The (new) prescription just given may not provide the best numerical algorithm for finding the pattern functions, but it does prove that a linear-transform inverse of Eq.(3.9) exists. This is written in full as (in the notation of [114])

$$\rho_{mn} = \frac{1}{2\pi} \int_{-\pi}^{\pi} d\theta \int_{-\infty}^{\infty} dq\, \exp[i(m-n)\theta]\, M_{mn}(q)\, \Pr(q,\theta) \quad , \quad (3.14)$$

where the functions $M_{mn}(q)$ are the elements of the dual basis, and are often called *pattern functions* and given in the form $f_{mn}(q) = (1/\pi) M_{mn}(q)$ [89]. Efficient algorithms for their numerical construction have been given [1, 89]. Several examples of the pattern functions are given in in the chapter by D'Ariano, Paris and Sacchi of this volume.

In experiments the continuous integral over phase in Eq.(3.14) is replaced by a sum over discrete, equally spaced phase values, $\theta_k$,

$$\rho^{(d)}_{mn} = \frac{1}{d} \sum_{k=1}^{d} \int_{-\infty}^{\infty} dq \, \exp[i(m-n)\theta_k] \, M_{mn}(q) \, \Pr(q, \theta_k) \qquad . \qquad (3.15)$$

In general, one would think the larger the number of phases $d$, the better. An analysis by Leonhardt and Munroe determined the minimum number of LO phases needed for a faithful state reconstruction [114]. They proved that if the field is known a priori to contain at most $\tilde{n}$ photons, then the number of LO phases needed equals simply $\tilde{n}+1$, the dimension of the Hilbert space containing the state. Equation (1.4) shows that the quadrature distribution contains oscillations as a function of phase that are determined by the occupied photon-number states; there is no further information to be had by using more phase values than $\tilde{n}+1$ since those higher phase "frequencies" would be aliased to lower frequencies, as is familiar in the Nyquist/ Shannon sampling concept [115]. Leonhardt and Munroe also show how to estimate the error incurred by choosing too few phases in the case that the maximum photon number is not known ahead of time.

A beautiful demonstration of the reconstruction of a density matrix in the number basis was carried out by Schiller et al. [42], and a comprehensive and very instructive report was given by Breitenbach [93]. A quadrature-squeezed field produced by optical parametric oscillation was detected by BHD to produce quadrature histograms for 128 phase values. The density matrix elements up to $n = 6$ were reconstructed using Eq.(3.15). The diagonal elements, or probabilities, which are plotted in Fig. 7, show clearly that even photon numbers (0, 2, 4) predominate, since photon-pair production is the origin of the quadrature-squeezed vacuum field. This was the first observation of the even-odd-number oscillations of squeezed light, an effect that could not be measured by using direct photon-number detection due to a lack (at that time) of photon-number-discriminating detectors. This shows one of the unique capabilities of the indirect measurement idea. Even stronger number oscillations were reconstructed from the same measured data by using non-deterministic, least-squares methods [93].

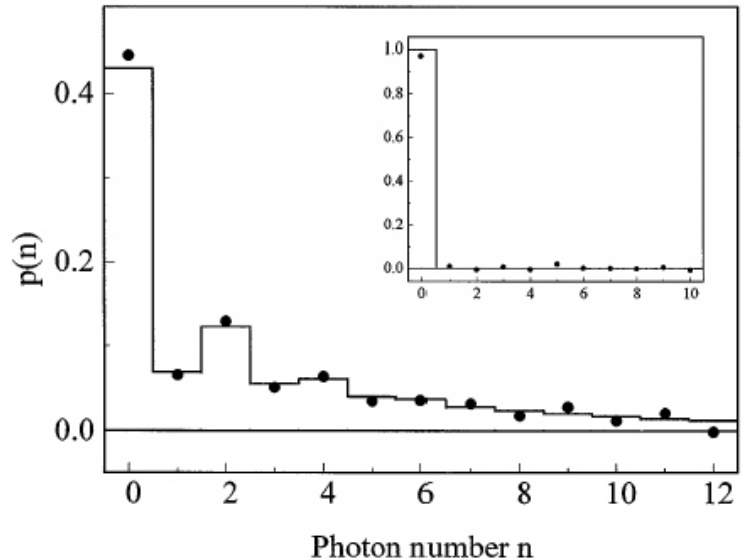

Fig. 7. Photon number distribution of squeezed vacuum and vacuum (insert). Continuous curves are theory. [42]

## 3.c Ultrafast Photon Number Sampling using Phase-Averaged BHD

An especially useful application of the pattern function idea is the indirect measurement of photon-number statistics by the use of a phase-random or phase-swept local oscillator (LO) [39, 116]. It is clear from Eq. (3.14) that the photon-number probability $\rho_{nn}$ is given by

$$\rho_{nn} = \int_{-\infty}^{\infty} d\xi \, M_{nn}(\xi) \, \overline{\Pr(\xi)} \quad , \tag{3.16}$$

where the phase-averaged quadrature distribution is

$$\overline{\Pr(\xi)} = \frac{1}{2\pi} \int_{-\pi}^{\pi} d\theta \, \Pr(q = \xi, \theta) \quad , \tag{3.17}$$

where $\xi$ represents the phase-independent quadrature variable. This means that in an experiment, only a *single* quadrature distribution needs to be measured, while randomizing or sweeping the phase uniformly in the interval $-\pi$ to $\pi$. This greatly reduces the amount of data needed to obtain the number statistics using OHT and also removes artifacts that may arise from the discrete stepping of the phase as is usually done in OHT.

The diagonal pattern (or sampling) functions are generated numerically most efficiently using the algorithm in [89]. Several examples are shown Fig. 8.

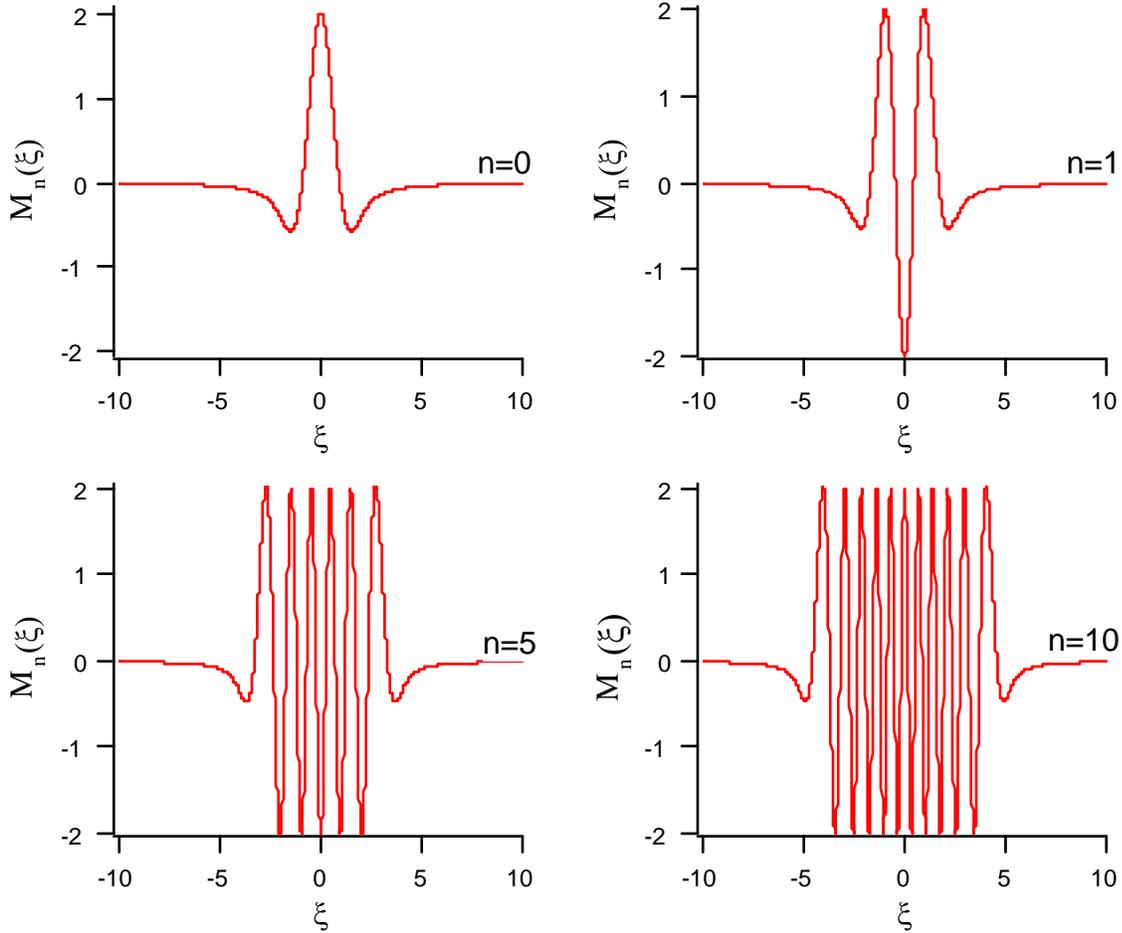

Fig. 8. Sampling functions $M_{nn}(\xi)$ versus $\xi$ for $n = 0, 1, 5,$ and $10$. [62]

An interesting feature of the functions $M_{nn}(\xi)$ is that their maxima and minima correspond to the maxima and minima of the quadrature distribution for a Fock state $\bar{P}(\xi) = \psi_n^*(\xi)\psi_n(\xi)$, where $\psi_n(\xi)$ is the wave function in the Fock basis.

This leads to the interesting and useful point that in many cases the best way to measure the statistical properties of photon numbers is to *not* measure photon number directly. Rather measure the quadrature-field amplitudes and infer the photon statistics. (See also the discussion in Sec. 1.) The latter has the advantages of high-quantum-efficiency detection (approaching 100%), good discrimination between probabilities of zero, one, two, etc. photon numbers, tens of fs time resolution, temporal and spatial mode selectivity. A complementary disadvantage is that if the average number of photons in the selected spatial-temporal mode is far less than unity, then the amount of signal averaging required to reconstruct the field and number statistics can make the technique impractical. Nevertheless, even in the case of photon numbers much less than one, the mean photon number (if not its statistics) in the detected mode can be measured using OHT if high pulse rates (> 1 MHz) are used.

Another drawback to the indirect measurement method is that it may lead to a lower precision of measurement than a direct method (if one is available) [117, 118]. This arises because a tomographic measurement is capable of yielding information about all of the system variables and, consistent with the uncertainty principle, one would expect a degrading of the precision in estimating a given observable's value from a finite data set. On the other hand, for determining distributions of some variables, such as photon number on fs time scales (see below), or the London/ Pegg-Barnett phase (see above), there are no direct measurement techniques known. Then indirect measurement provides a useful path, as illustrated further in the following.

Perhaps the most interesting properties of the photon number of a given mode are its mean value and its fourth-order statistics. A nice property of *phase-averaged homodyne detection* is its ability to extract these quantities from the quadrature data without the need to first reconstruct the quantum state or even the photon number distribution. Munroe et al. showed that the statistical moments of photon number can be computed directly from the raw quadrature data [39, 62].

For example, the mean photon number in the detected signal mode is proportional to the expectation value of the square of the quadrature amplitude, averaged over all phase values,

$$\langle \hat{n} \rangle = \langle \hat{a}^\dagger \hat{a} \rangle = \frac{1}{2\pi} \int_0^{2\pi} \langle \hat{q}_\theta^2 \rangle \, d\theta - \frac{1}{2} = \langle\langle q^2 \rangle\rangle - \frac{1}{2} \quad . \tag{3.18}$$

The double bracket in the last term indicates an experimental average over LO phase and measured quadrature values $q$ (i.e., $\xi$). The subtraction of 1/2 removes the vacuum (zero-point) contribution.

Munroe also derived the minimum number $N_{min}$ of measurements needed to reliably determine the mean photon number [62],

$$N_{min} = \frac{3}{2} \frac{\langle \hat{n}^2 \rangle + \langle \hat{n} \rangle + 1/2}{\langle \hat{n} \rangle^2} \quad . \tag{3.19}$$

For example, a field that is coherent or thermal with mean number $10^{-3}$ would require 750,000 measurements to reach a signal-to-noise ratio equal to one for the mean photon number. Munroe compared this to a similar quantity for standard photon counting in the presence of a background count, $N_{min}^{PC} = [\langle \hat{n}^2 \rangle + 2\langle \hat{n} \rangle\langle \hat{n}_B \rangle + \langle \hat{n}_B^2 \rangle]/\langle \hat{n} \rangle^2$ [119]. He pointed out that the two nearly coincide in the case that the background has a mean value of 1/2 photon. This arises because BHD detects the vacuum field, showing why standard counting can be superior to BHD for very weak signals in the absence of significant background.

Formulas for higher-order moments of the photon-number distribution can be calculated similarly to Eq. (3.18), using Richter's formula for the number factorial moment [120],

$$\langle n^{(r)} \rangle = \sum_{n=0}^{\infty} [n(n-1)...(n-r+1)] \, p(n) = \langle (\hat{a}^{\dagger})^r (\hat{a})^r \rangle$$

$$= \frac{(r!)^2}{2^r (2r)!} \int_0^{2\pi} \frac{d\theta}{2\pi} \langle H_{2r}(\hat{q}_\theta) \rangle \quad , \quad (3.20)$$

where $H_j(x)$ is the Hermite polynomial. An important quantity that can be derived from this is the "second-order coherence," defined as the normalized, normally ordered number-squared,

$$g^{(2)}(t,t) = \frac{\langle :\hat{n}^2: \rangle}{\langle \hat{n} \rangle^2} = \frac{\langle \hat{n}^2 \rangle - \langle \hat{n} \rangle}{\langle \hat{n} \rangle^2} \quad , \quad (3.21)$$

where the time argument indicates that the sampling takes place using an LO pulse centered at time *t*. This quantity is computed from the data using [62, 120]

$$g^{(2)}(t,t) = \frac{(2/3)\langle\langle q^4 \rangle\rangle - 2\langle\langle q^2 \rangle\rangle + 1/2}{\langle\langle q^2 \rangle\rangle^2 - \langle\langle q^2 \rangle\rangle + 1/4} \quad . \quad (3.22)$$

Munroe has also derived a scheme for computing the statistical uncertainties of such moments directly from the raw data, making it possible to put error bars on the measured values [62, 89]. For example, the variance in the measured mean photon number $\langle n \rangle$ is estimated by $N^{-1} \langle\langle \xi^4 \rangle\rangle$, where *N* is the total number of pulses sampled. And the variance in the measured photon number probability *p(n)* is estimated by $N^{-1} \langle\langle M_{nn}^2(\xi) \rangle\rangle \leq 4/N$.

In the case that the LO has the form of an ultrashort pulse, BHD provides a fast time gating or windowing capability, as pointed out in connection with Eq.(2.13) above. This was first demonstrated by Munroe et al. for the 5-ns signal pulse from a semiconductor diode laser below threshold [39]. The diode laser was pumped electrically by a voltage pulse synchronized to the mode-locked pulse train from the fs laser that served as the LO. This is the first example of OHT being applied to a signal that was not optically pumped by light derived from the LO laser. In this case the LO phase needs not be swept, since the signal source is intrinsically phase random.

Munroe also carried out detailed studies of number statistics of light emitted by a traveling-wave semiconductor amplifier, or super luminescent diode (SLD) [62]. This is a diode with gain but virtually no cavity feedback. Amplified spontaneous emission (ASE) from amplifiers plays an important role as noise in optical communication systems. When electrically pulsed synchronously with the LO clock, the SLD emits ASE as a 5-ns pulse which is sampled using the random-phase BHD technique. The SLD could be operated either in a single-pass configuration, with no feedback, or with one-sided feedback (but no cavity) being provided by a diffraction grating. After 10,000 samples were collected using a 150-fs LO pulse that could be varied in its delay, the mean photon number and second-order coherence were computed as in Eqs. (3.18) and (3.22). These are shown in Fig. 9.

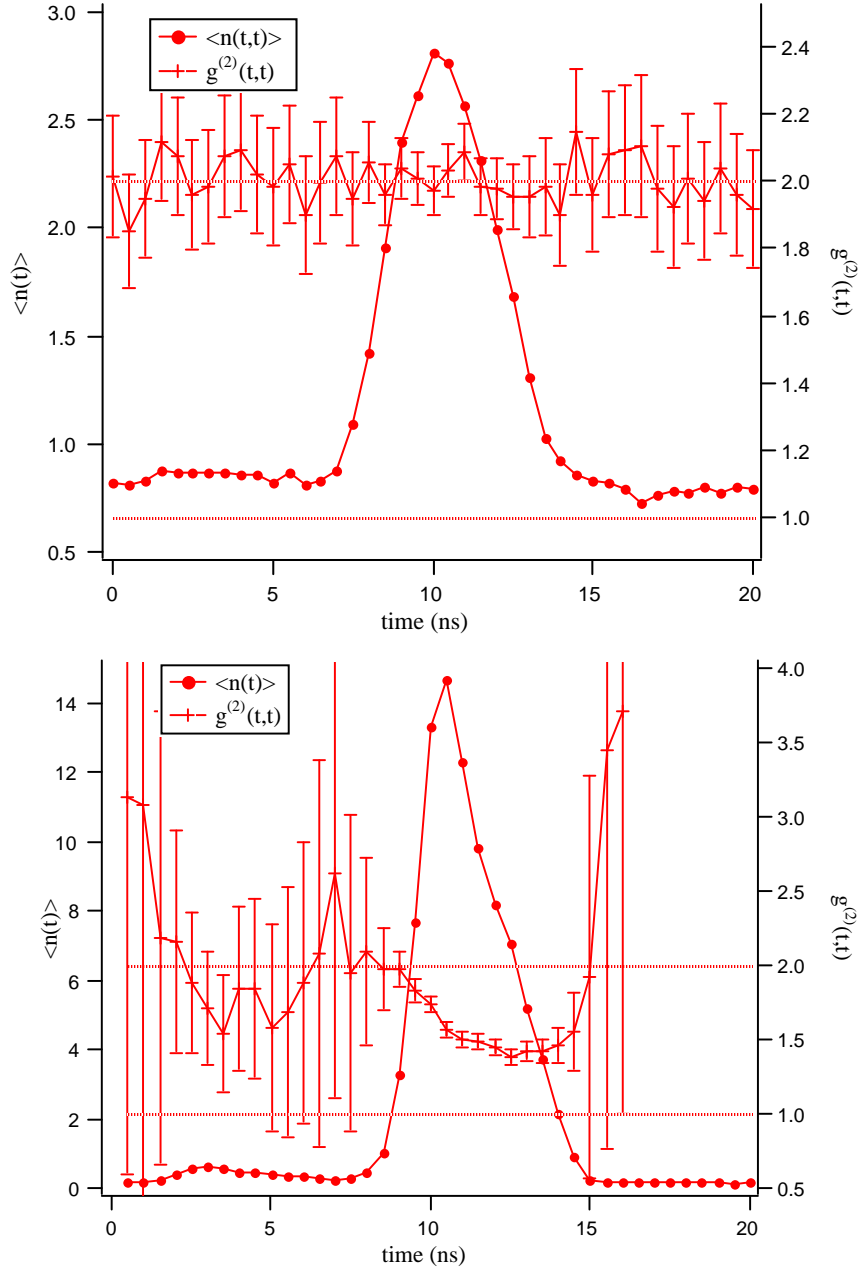

Fig. 9. Average photon number $\langle n(t)\rangle$ versus time (•) and the second-order coherence $g^{(2)}(t,t)$ versus time (+) for (a) the SLD in the single-pass configuration, and (b) in the double-pass with grating configuration [62].

A value of $g^{(2)}(t,t) = 2$ corresponds to thermal light, i.e. light produced primarily by spontaneous emission, and a value of $g^{(2)}(t,t) = 1$ corresponds to light with Poisson statistics, i.e., light produced by stimulated emission in the presence of gain saturation. Figure 10 shows plots of the photon number distribution, with error bars, of the SLD emission determined from the measured quadrature distributions.

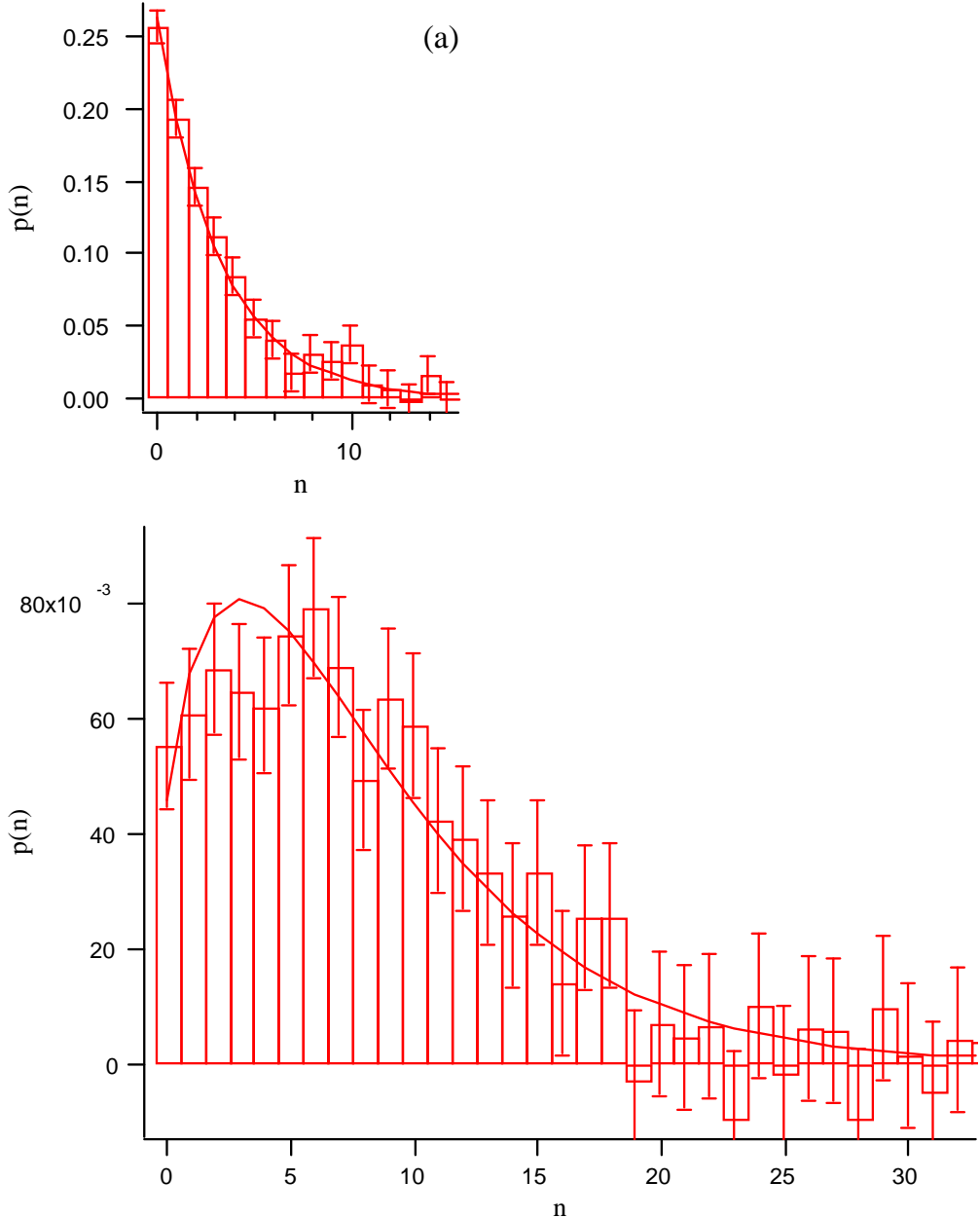

Fig. 10. Measured photon number distributions at the pulse peaks for SLD data in Fig. 9, (a) single-pass and (b) double-pass with grating. The solid curves are (a) Bose-Einstein distribution, and (b) negative-binomial distribution. From [62].

As we have seen, BHD with a pulsed LO provides a powerful technique for ultrafast time-gated detection of the signal field rather than its intensity. It has recently been developed into a practical scheme called *Linear Optical Sampling*, useful in, for example, testing fast optical telecommunications hardware [41, 48, 94]. It is useful to ask what fundamentally limits the time resolution of this sampling technique. It is clear from the above discussion that this resolution can be as short as the duration of the LO pulse. But it is easy to see that there are special cases in which the time resolution can be far better than would appear to be set by the LO duration itself. Consider that the LO temporal mode can be written as $\hat{\underline{\Phi}}_L^{(+)}(\underline{x},0,t) = i\sqrt{c}\ \hat{c}_L \underline{v}_L(\underline{x}) f_L(t-\tau)$, that is, a product of a normalized transverse spatial part and a normalized temporal part $f_L(t-\tau)$ that is delayed by a variable time $\tau$. (This can be accomplished experimentally in a given plane, where the detector is located, while at any other plane the field will in general not factor.) Then the difference number in Eq.(2.8) can be written

$$\hat{N}_-(\tau) = -i\sqrt{c}\, \alpha_L^* \int_0^T dt\, f_L^*(t-\tau)\, \phi_S(t) \; + \; h.c. \qquad , \qquad (3.23)$$

where the signal-field time dependence is given by

$$\phi_S(t) = \int_{Det} d^2x\, \underline{v}_L^*(\underline{x}) \cdot \hat{\underline{\Phi}}_S^{(+)}(\underline{x},0,t) \qquad . \qquad (3.24)$$

This can be written in the Fourier domain as

$$\hat{N}_-(\tau) = -i\sqrt{c}\, \alpha_L^* \int_{-\infty}^{\infty} \frac{d\omega}{2\pi} e^{-i\omega\tau}\, \tilde{f}_L^*(\omega)\, \tilde{\phi}_S(\omega) \; + \; h.c. \qquad . \qquad (3.25)$$

Now consider the special case that the signal is band-limited with bandwidth $B$, that is, it equals zero outside of a spectral interval $[\nu - B/2, \nu + B/2]$ around a central frequency $\nu$. Further assume that the LO field is constant throughout this interval. Then we have (exactly) [94],

$$\hat{N}_-(\tau) = -i\sqrt{c}\, \alpha_L^* \tilde{f}_L^*(\nu) \int_{\nu-B/2}^{\nu+B/2} \frac{d\omega}{2\pi} e^{-i\omega\tau}\, \tilde{\phi}_S(\omega) \; + \; h.c.$$

(3.26)

$$= -i\sqrt{c}\, \alpha_L^* \tilde{f}_L^*(\nu)\, \phi_S(\tau) \; + \; h.c.$$

This remarkable result shows that, by using an LO pulse having a finite duration and a Fourier spectrum $\tilde{f}_L(\nu)$ that is constant in the interval containing the signal spectrum, a band-limited signal field can be sampled exactly, with a resolution that is not degraded by the finite duration of the LO pulse. An example of such an LO pulse is $f_L(t) \propto (1/t)\sin(Bt/2)$. This result is closely related to the Whittaker-Shannon sampling theorem [115], which is used here in an inverse manner. Our result is most useful for measuring repetitive signals whose form is the same from pulse to pulse.

### 3.d Dual-LO BHD and Two-Mode (or Two-Time) Tomography

There is no reason that the LO pulse needs to have a simple single-pulse form. If the LO instead is comprised of two well isolated short pulses (with the same or differing spatial forms), then we may choose to view it as a linear superposition of two distinct wave packet modes. (The same idea applies to two polarization modes; see below.) In general the LO field is written as [9, 121-123]

$$\hat{\underline{\Phi}}_L^{(+)}(\underline{r},t) = i\sqrt{c}\, |\alpha_L| \exp(i\theta)\left[\underline{v}_1(\underline{r},t)\cos\alpha + \underline{v}_2(\underline{r},t)\exp(-i\zeta)\sin\alpha\right] \qquad , \qquad (3.27)$$

where $\alpha$ and $\zeta$ are parameters that can be varied, and the two mode functions are orthonormal. When this is inserted into Eq.(2.14) we find that the detected quadrature corresponds to the mode operator

$$\hat{a} = \sum_k \hat{a}_k\, c\int_0^T dt \int_{Det} d^2x\; \underline{v}_k(\underline{x},0,t)$$
$$\times \left[\underline{v}^*_1(\underline{x},0,t)\cos\alpha + \underline{v}^*_2(\underline{x},0,t)\exp(i\zeta)\sin\alpha\right] \qquad (3.28)$$
$$= \hat{a}_1\cos\alpha + \hat{a}_2\exp(i\zeta)\sin\alpha \; ,$$

where $\hat{a}_1$ and $\hat{a}_2$ are the operators for the components of the signal field in each of the modes of interest. Using $\hat{a}_1 = (\hat{q}_1 + i\hat{p}_1)/\sqrt{2}$ and $\hat{a}_2 = (\hat{q}_2 + i\hat{p}_2)/\sqrt{2}$, we find that the measured quadrature $\hat{Q} = (\hat{a}e^{-i\theta} + \hat{a}^\dagger e^{i\theta})/2^{1/2}$ is

$$\hat{Q} = \cos(\alpha)[\hat{q}_1\cos\theta + \hat{p}_1\sin\theta] + \sin(\alpha)[\hat{q}_2\cos(\theta-\zeta) + \hat{p}_2\sin(\theta-\zeta)] \qquad . \qquad (3.29)$$

If we define $\theta - \zeta = \beta$ then the bracketed terms are recognized as the phase-dependent quadratures defined in Eq.(2.18) (one for each mode), so we can write

$$\hat{Q} = \cos(\alpha)\hat{q}_{1\theta} + \sin(\alpha)\hat{q}_{2\beta} \qquad . \qquad (3.30)$$

This shows that we can measure, in a single event, an arbitrary linear superposition of the quadratures of two modes using dual-LO BHD (but, of course, not the variable conjugate to $\hat{Q}$). The phases $\theta$ and $\beta$ of the two LO fields, as well as $\alpha$, are independently variable by the experimenter.

An intriguing application of the two-mode tomography concept is in the reconstruction of two-time statistics of a signal field [43, 124]. It should be pointed out that each subsystem is measured only once and then discarded, so measurement-induced dynamics cannot be measured by this technique – the system evolves as if no measurements were made. In this sense the situation is the same as in the well-known Hanbury Brown-Twiss correlation measurements, but in principle we can measure the complete two-time state evolution, not just the correlation of two particular observables. Then, from the reconstructed two-time (i.e., two-mode) state one could calculate the correlation function of any two variables at the two times selected.

It has been shown that it is possible to reconstruct the joint quantum state of the combined two-mode system by measuring probability histograms for the combined quadrature Q for many values of the parameters $\alpha$, $\theta$ and $\beta$ [121, 122]. Reconstruction in the number basis uses two-mode pattern functions [121].

If only the two-mode joint photon statistics are desired, then the two components of the combined LO field can have uniformly random phases, and a simpler two-mode pattern function can be used [121]. The theoretical scheme has also been generalized to the case of many modes [125].

A complete two-mode state reconstruction, which uses two-mode pattern functions, has not been performed in a laboratory to our knowledge, illustrating the challenge this poses regarding the amount of data required and the technical difficulty of controlling all parameters in the experiment well enough during the data collecting time [126]. Nevertheless, the technique has been applied to full reconstruction of two-mode photon-number statistics and correlations [57], as well as two-time number correlations on ps time scales [43]. Here we review the latter application.

The first experimental demonstration of ultrafast two-time number correlation measurements using dual-LO BHD was carried out by McAlister [43], who reconstructed the two-time second-order coherence defined (for a quasi-monochromatic field) as

$$g^{(2)}(t_1, t_2) = \frac{\langle :\hat{n}(t_1)\hat{n}(t_2): \rangle}{\langle \hat{n}(t_1) \rangle \langle \hat{n}(t_2) \rangle} \qquad , \qquad (3.31)$$

which generalizes Eq.(3.21). The signal source studied was a super luminescent laser diode (SLD, see above). While the LO was temporally synchronized with the 4-ns SLD pulse, there was no need for phase coherence between the signal and the LO since the signal field was intrinsically phase-random. In this special case the relative phase between the two LO pulses also is not important. The 150 fs LO pulses from a Ti:sapphire laser were split and, after a variable delay, recombined to make a dual-pulse LO in the form of Eq.(3.27) with adjustable $\alpha$, and random (or arbitrary) $\theta$ and $\beta$. The second-order coherence $g^{(2)}$ is reconstructed from combinations of second and fourth-order moments of Q obtained at three different values of $\alpha$: 0, $\pi/4$, and $\pi/2$. These three values correspond simply to a measurement of the first mode by itself, a measurement of the second mode by itself, and a measurement of an equal combination of both modes. Formulas are known also for estimating the statistical errors of the $g^{(2)}$ measurement [124].

In Fig. 11 we show the SLD data from the thesis of McAlister [109].

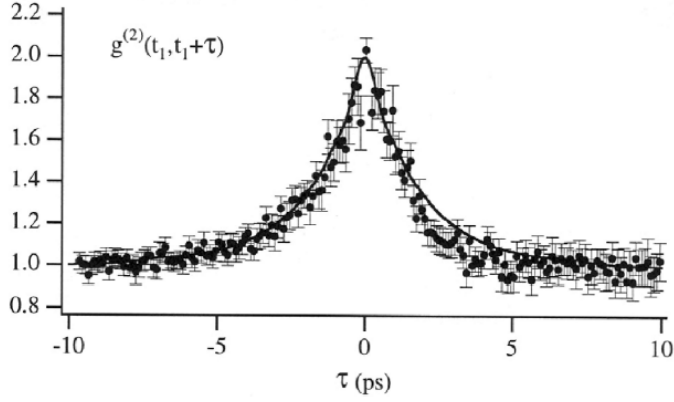

Fig. 11. Measured two-time photon number correlation $g^{(2)}(t_1, t_1 + \tau)$ from a pulsed SLD at low power. The solid curve is obtained from the Fourier transform of the measured optical spectrum of the source, under the assumption that the field is quasi-stationary and the statistics are thermal-like. [109]

### 3.e Optical Polarization Tomography

A dual-LO scheme for the purpose of Optical Polarization Tomography was proposed and analyzed in [127]. This can reconstruct the polarization state of a beam-like spatial mode comprised of two orthogonal polarization modes. Rather than requiring a full reconstruction of the two-mode state (density matrix), the measurement of the quantum state of polarization requires reconstruction only of a subset of the density matrix that we called the "polarization sector." This is defined to be the number-basis elements $_1\langle n_1|_2\langle n_2|\hat{\rho}|n'_1\rangle_1|n'_2\rangle_2$, restricted to $n_1 + n_2 = n'_1 + n'_2$, where $n_1$ and $n_2$ refer to the number of photons in each polarization. Knowledge of this portion of $\hat{\rho}$ is sufficient to calculate any statistical moments of the polarization Stokes operators, which are defined by $\hat{J}_1 = (\hat{a}_1^\dagger \hat{a}_1 - \hat{a}_2^\dagger \hat{a}_2)/2$, $\hat{J}_2 = (\hat{a}_1^\dagger \hat{a}_2 + \hat{a}_2^\dagger \hat{a}_1)/2$, $\hat{J}_3 = (\hat{a}_1^\dagger \hat{a}_2 - \hat{a}_2^\dagger \hat{a}_1)/2i$. In terms of the angular momentum eigenstates $|j,m\rangle$ of $\hat{J}^2$ and $\hat{J}_1$, the polarization sector corresponds to the elements $\langle j,m|\hat{\rho}|j,m'\rangle$ for all $j, m, m'$.

The SU(2) group of unitary transformations on the $\hat{J}_i$ operators is equivalent to the general two-mode transformation of the mode operators $\hat{a}_1$, $\hat{a}_2$ to give new operators $\hat{a}_3$, $\hat{a}_4$,

$$\hat{a}_3 = \hat{a}_1 \cos(\gamma/2) + \hat{a}_2 \exp(i\zeta)\sin(\gamma/2)$$
$$\hat{a}_4 = -\hat{a}_1 \sin(\gamma/2) + \hat{a}_2 \exp(i\zeta)\cos(\gamma/2) \quad , \quad (3.32)$$

which preserves the commutator, $[\hat{a}_3, \hat{a}_4^\dagger] = 0$. Note that, with the identification $\gamma/2 = \alpha$, the operator $\hat{a}_3$ is the same as $\hat{a}$ defined in Eq.(3.28) and measured using dual-LO BHD. On the Poincare' sphere representation of the $\hat{J}_i$ operators, the $\hat{J}$ vector is rotated by angles $\gamma$ and $\zeta$ when the $\hat{a}_i$ operators undergo the transformation Eq.(3.32), as illustrated in Fig. 12.

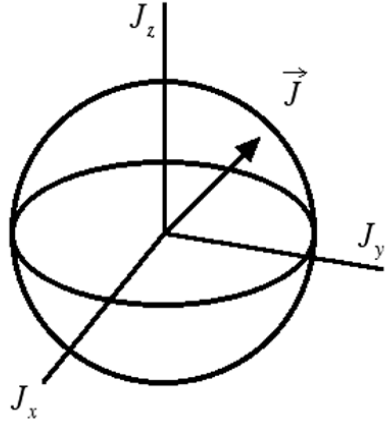

Fig. 12. Poincare' sphere representation of the optical polarization state.

It was shown in [127] that full reconstruction of the polarization sector can be accomplished by using a dual-polarization-mode LO of the form Eq.(3.27), with the LO components having identical spatial-temporal forms but orthogonal polarizations. In this case the overall (common-mode) phase $\theta$ can be randomized or swept uniformly over $0 - 2\pi$. This arises because polarization depends only on the relative phase between two modes, and significantly reduces the amount of data collection required for a reconstruction, compared to a full two-mode state reconstruction.

As shown by Bushev et al., a complete characterization of the quantum state of optical polarization (but not the complete two-mode state) can also be obtained by using direct photoelectric detection rather than homodyne detection [128]. One does this by measuring joint photoelectron statistics of the photon numbers of the two polarization modes following many different polarization-state transformations. From measured histograms the state is extracted in the form of a polarization Wigner distribution on the Poincare' sphere, whose marginal distributions give probabilities for a general polarization (Stokes) variable, from which arbitrary moments can be computed [129, 130]. Experimental results for polarization-squeezed light were reported [128].

If only first and second moments of the Stokes operators are desired, a simpler scheme can be used, in which only three mode transformations are needed before photoelectron statistics are collected [131]. This method is employed in a modified manner in the experiments discussed below in Sec. 3.g.

A smaller polarization sector can be used if it is known *a priori* (or determined by post selection) that the number of photons in a given set of modes is limited to a fixed number. For example, the recent experiments by White et al. [21], reconstructing the quantum state of polarization of pairs of photons, correspond to measuring a small subset of a polarization sector for a four-mode density matrix (two polarization modes for each photon). See the chapter in this volume by Altepeter, James and Kwiat.

### 3.f  Two-Mode Tomography by Generalized Rotations in Phase Space (GRIPS)

It was pointed out in [132] and Richter [133] that in some cases it may be easier to obtain tomographic state information about a pair of modes by using a fixed, single-mode (single-wave packet) LO field and implementing a two-mode transformation on the signal field [rather than on the LO field as in Eq.(3.27)]. The most general unitary transformation [SU(2)] on a pair of mode operators $\hat{a}_1$, $\hat{a}_2$ defines new operators $\hat{a}_3$, $\hat{a}_4$ as given in Eq.(3.32) above. The idea is to measure quadrature statistics for operator $\hat{a}_3$ under many different transformation conditions. This yields enough information to reconstruct the state. Note again that the operator $\hat{a}_3$ is the same as $\hat{a}$ defined in Eq.(3.28).

For our purposes only $\hat{a}_3$ need be measured. This is done using a linear-optical device, consisting of a pair of controllable birefringent phase retarders followed by a polarizing beam splitter (PBS) to separate the resulting modes $\hat{a}_3$ & $\hat{a}_4$. Mode $\hat{a}_3$ is created such that it has the same polarization as the LO. It and the LO enter the DC balanced-homodyne detector, which linearly combines (interferes) mode $\hat{a}_3$ and the LO mode having phase $\theta$. After subtraction and scaling of the photodetectors' signals, the quantity measured on a single trial is the "combined quadrature amplitude,"

$$Q(\gamma/2, \theta, \zeta) = [a_3 \exp(-i\theta) + a_3^\dagger \exp(i\theta)]/2^{1/2}$$
$$= \cos(\gamma/2)\, q_{1,\theta} + \sin(\gamma/2)\, q_{2,\theta-\zeta} \quad . \tag{3.33}$$

This shows that by varying the parameters $\gamma$, $\theta$ and $\zeta$ the experimenter can measure the combined quadrature amplitude of the two-mode signal following a rotation by arbitrary angles on the Poincare' sphere. This provides a tomographically complete set of observables, whose statistical characterization allows reconstruction of the full quantum state by using the two-mode pattern functions mentioned in Sec. 3.d above.

As before, it is possible to obtain directly certain field moments or number moments by measuring quadrature data at only a small set of well chosen values of $\gamma$, $\theta$ and $\zeta$ [43]. For a summary of various moment formulas, see [109].

**3.g  Ultrafast Optical Polarization Sampling by GRIPS Tomography**

The statistical correlation between two orthogonal polarization modes (*i* and *j*) with photon numbers $n_i(t_1)$ and $n_j(t_2)$ at two times $t_1$ and $t_2$ is characterized by the normalized two-mode, two-time, second-order coherence (correlation) function

$$g_{i,j}^{(2)}(t_1, t_2) = \frac{\langle :\hat{n}_i(t_1)\,\hat{n}_j(t_2): \rangle}{\langle \hat{n}_i(t_1)\rangle \langle \hat{n}_j(t_2)\rangle} \quad . \tag{3.34}$$

(The normal ordering is unnecessary unless $i = j$.) A value $g_{i,j}^{(2)} = 1$ indicates uncorrelated fluctuations in $n_i(t_1)$ and $n_j(t_2)$, and a value above (below) 1 indicates positive (negative) correlations. Such a quantity can be measured by the sampling schemes discussed above–either the dual-LO or the GRIPS scheme.

A version of the GRIPS scheme allows two-mode correlations to be measured without any assumption about the photon number. The analysis of optical polarization follows an earlier proposal by Karassiov and Masalov (KM) [131], and was implemented by Blansett, et al., for the purpose of studying ultrafast polarization dynamics of 30-ps pulses from a vertical-cavity surface-emitting semiconductor laser (VCSEL) [50, 134].

In the resulting "two-time optical polarization sampling method," shown in Fig.13, Blansett employed the idea of KM to measure the signal beam separately in three distinct polarization bases–R/L, H/V, or +45/-45, where R (L) means right (left) circular, H (V) means horizontal (vertical), and vertical +45 (-45) means + (-) 45 degree linear polarizations. To accomplish this basis resolution, the signal beam emitted by the VCSEL may pass through one or two waveplates (or it may pass through no waveplates.) The first is a quarter-wave plate (QWP),which converts R circularly polarized light into V linearly polarized light, and L polarized light into H polarized light. The second is a half-wave plate (HWP) which converts +45 deg polarization into V polarization, and -45 deg linear into H. KM pointed out that measuring means and fluctuations of intensity separately in all three bases allows one to characterize fully the polarization statistics up to second order, that is, the first and second moments of the Stokes operators. Blansett's generalized scheme also allows measuring the correlations between orthogonally polarized modes, which cannot be extracted from moments of the Stokes operators alone [134].

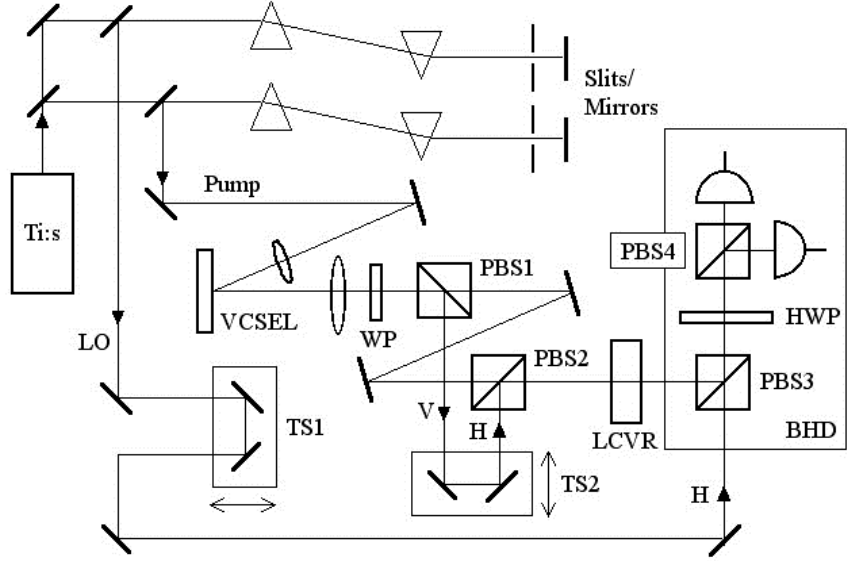

Fig. 13. Setup for sampling scheme for one- or two-time optical polarization correlations. TS–translation stages; WP–wave plate (optional); HWP–half-wave plates; LCVR–liquid-crystal variable-retarder; BHD–balanced-homodyne detector. [134]

Whereas in the KM scheme fast photodetectors would ordinarily be used to make time-resolved measurements of intensity in the three bases, Blansett adopted ultrafast, phase-averaged BHD (as described above) to sample these statistics on sub-ps time scales. This is accomplished by splitting the signal after the WP using a polarizing beam splitter (PBS1) into V and H-polarized beams. These two are recombined at PBS2 with near-zero (few fs) time delay after the V component has had its phase shifted by a movable mirror (TS2) whose purpose is to sweep uniformly over $0 - 2\pi$ during the data acquisition of quadrature histograms (typically requiring 20,000 pulses). The LO overall phase is also uniformly swept by moving translation stage TS1. The phase sweeping allows the use of the simpler two-mode tomographic reconstruction scheme mentioned above.

Note that for this experiment the VCSEL emission has a longer wavelength than does the laser pulse that pumps the VCSEL. The pump pulse and the LO pulse, which must have the same wavelength as the VCSEL signal, are derived from the wide-band Ti:sapphire pulse by use of prism-slit-based spectral filters, and both had durations of 300 fs.

In order to select the two-mode quadrature to be measured in the BHD, a device (liquid-crystal variable-phase retarder, LCVR) is used as a computer-controllable half-wave plate to rotate the linear polarizations exiting PBS2 by 90 deg, 45 deg, or 0 deg. Then PBS3 reflects only V polarization into the path of the H-polarized LO beam. Finally, a HWP rotates by 45 deg and PBS4 projects out + and – linear superpositions of LO and combined-signal beam. The choice of rotation angle induced by the LCVR sets the value of $\gamma$ (0, $\pi/2$, or $\pi$), in the combined quadrature

$$Q = \cos(\gamma/2)\, q_{1,\theta} + \sin(\gamma/2)\, q_{2,\theta-\zeta} \qquad , \qquad (3.35)$$

and the phases $\theta$ and $\zeta$ are swept (and averaged over) in the manner described above. The moments are calculated using the formulas given in [109]. Statistical error bars are calculated according to formulas in [124].

In the first measurements shown in Fig.14, Blansett used a near-zero time delay in TS2 so the scheme measures a one-time, two-polarization correlation and corresponds to the GRIPS method described above. For the low-temperature VCSEL emission in Fig.14a, the R and L emission modes show $g^{(2)}_{RR} = g^{(2)}_{RR} \cong 2$ at early and late times, corresponding to spontaneous emission, and $g^{(2)}_{RR} = g^{(2)}_{RR} \cong 1.5$ at the emission peak, indicating lasing or stimulated emission, with photon statistics tending toward a Poisson number distribution. $g^{(2)}_{RL} \cong 1$ shows that the R and L modes emit in uncorrelated fashion. This is in contrast to results obtained with a room-temperature VCSEL, Fig 14(b), which shows anticorrelated R and L intensities $g^{(2)}_{RL} < 1$ [equation added]. Theoretical modeling reveals that this anticorrelation is caused by a higher spin-flip rate at the higher temperature, leading to gain competition between modes. This new measurement capability allows unprecedented time resolving of such lasing dynamics and statistics, described in detail in [50].

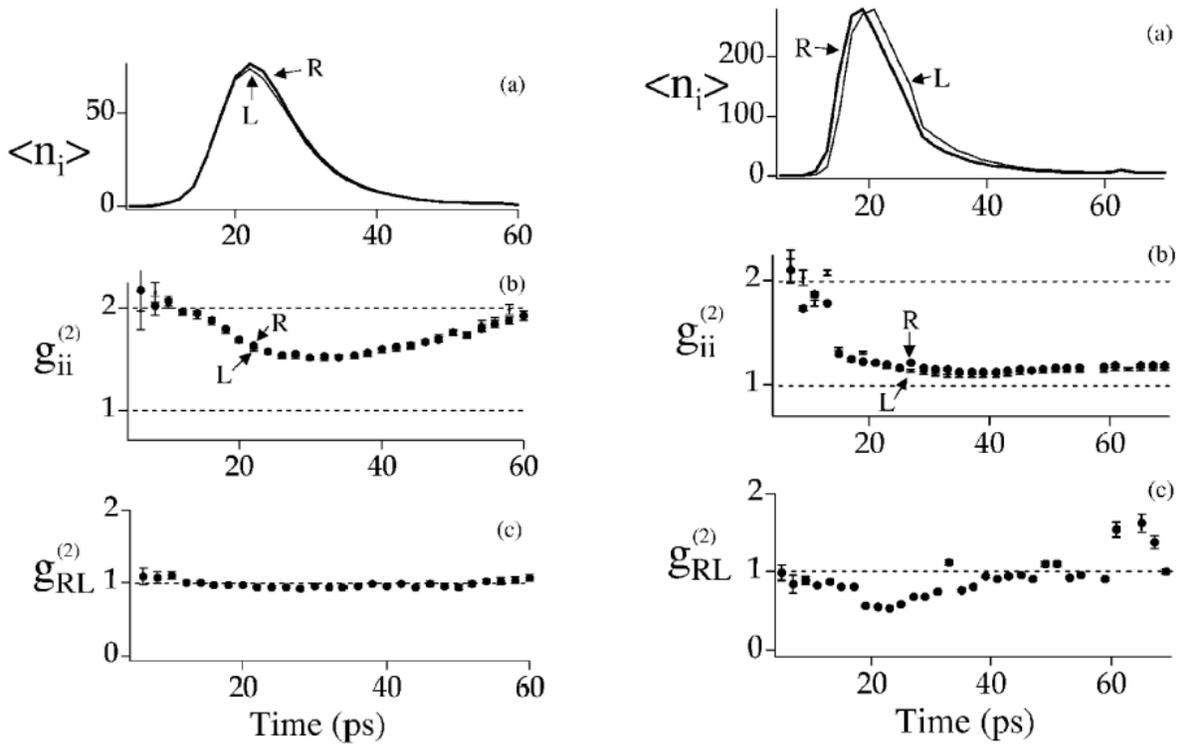

Fig. 14. One-time optical polarization mean photon number <n> and correlations, with H-polarized pump, for (a) low-temperature (10 K) VCSEL and (b) room-temperature VCSEL [134]

In the second set of measurements (not shown) Blansett used a non-zero time delay in TS2, so the scheme measures a two-time, two-polarization correlation.

### 3.h Simultaneous Time and Frequency Measurement

An important property of the dc-balanced homodyne technique, not pointed out previously, is that it provides spectral as well as temporal information about the signal field. This arises because if the LO field is frequency-tuned [by harmonically varying the function $f_L(t)$] away from the spectral region of the signal, the integral in Eq. (1.15) defining the mode-matched amplitude $\hat{a}$ will decrease. To analyze this, define the LO temporal function to be

$$f_L(t) = e^{-i\omega_L t} h_L(t - t_L) \quad , \quad (3.36)$$

where $\omega_L$ is the LO's center frequency and $h_L(t - t_L)$ is a real function with maximum at $t = t_L$. One way to achieve this would be by generating an ultrashort pulse (e.g., 30fs) and passing it through a tunable bandpass filter, followed by a time-delay. Using the reconstructed joint statistics of quadrature amplitudes, the mean number of mode-matched signal photons $\bar{N}_{MMS}(\omega_L, t_L) = \langle \hat{a}^\dagger \hat{a} \rangle$ can be determined for a set of $\omega_L$ and $t_L$ values. In the semiclassical case, this quantity, for a given LO center frequency $\omega_L$, is equal to

$$\bar{N}_{MMS}(\omega_L, t_L) = \left\langle \left| \int_0^T dt\, e^{i\omega_L t} h_L(t - t_L) \phi^{(+)}(t) \right|^2 \right\rangle \quad , \quad (3.37)$$

where $\phi^{(+)}(t)$ is the spatial-mode-matched signal,

$$\phi^{(+)}(t) = \int_{Det} d^2 s\, u_L^*(\underline{s}) \Phi_S^{(+)}(\underline{s}, t) \quad , \quad (3.38)$$

and the brackets indicate an average over multimode coherent-state amplitudes. Equation (3.37) is identical to the general form for time-dependent spectra [135], with $h_L(t - t_L)$ acting as a time-gate function. For example, it can be put into the same form as appearing in the "time-dependent physical spectrum" [136-138] if we specialize to

$$h_L(t - t_L) = \begin{cases} 0, & t > t_L \\ e^{-\gamma(t - t_L)}, & t < t_L \end{cases} \quad . \quad (3.39)$$

Thus by scanning the LO center frequency $\omega_L$ and arrival time $t_L$ independently, and measuring $\bar{N}_{MMS}(\omega_L, t_L)$, one obtains both time and frequency information, within the usual time-frequency bandwidth limitations. This method provides an alternative to the nonlinear optical upconversion technique which has been used to measure time-frequency information for light emitted by vibrational molecular wave packets [22].

## 4. Experimental Techniques

Detection circuits for balanced detection fall into two major classes: radio frequency (RF) detection and charge-sensitive whole-pulse detection (which we refer to as DC detection). Recently array detection has been used for quantum state measurements [31-33]; array detection is a generalized version of DC detection. We will discuss each of these detection technologies, with an emphasis on DC detection.

We also note that quantum-state measurement of electromagnetic fields has been performed with single-photon counting detectors [19], and by probing an optical field in a cavity with Rydberg atoms [56]. These state measurement techniques are very different from OHT, and we will not discuss them in detail.

### 4.a DC Detection

DC detection was used by Smithey et al. in the first demonstration of OHT [16]. It has been the workhorse for our experiments ever since, and it is the system we have the most familiarity with. To our knowledge, Guéna et al. were the first to operate such a detector at the shot-noise level (SNL) [100], while the first detection of nonclassical light using this technology was performed by Smithey et al. [99]. Other implementations of this technology have been performed by Hansen et al. [51] and Zavatta et al. [52].

In this detection technique a short pulse of light (usually much shorter than the response time of the detector) is incident on a photodiode and produces a charge pulse. On average, the number of photoelectrons in the output pulse is equal to the number of incident photons times the quantum efficiency $\eta$ of the photodiode. The goal is to measure the number of photoelectrons produced by each pulse as precisely as possible.

In order to be digitized by an analog-to-digital converter (ADC) and stored in a computer, a charge pulse must be converted to a voltage pulse and then amplified. A charge-sensitive preamplifier (also known as a charge integrator) followed by pulse-shaping electronics does this. The electronics used are common in nuclear spectroscopy or x-ray detection. We refer to this detection technique as DC because the preamplifiers integrate the total charge per pulse, and we sample the data synchronously with the pulse repetition frequency. Every photoelectron entering the amplifier is counted; we do not measure the current within some bandwidth about a nonzero offset frequency.

A schematic of the electronics is shown in Fig. 15. The photodiode is reverse biased with a DC voltage; the input coupling capacitor blocks this DC bias and passes the short (a few ns) current pulse produced when an optical pulse hits the photodiode. The charge integrator integrates the input charge, and puts out a voltage pulse with a rise time on the order of a few nanoseconds, and a fall time on the order of 100's of microseconds. The peak voltage of this output pulse is proportional to the total input charge; the fall time is determined by the time constant of the integrator $R_iC_i$. The pulse shaper (also known as a spectroscopy amplifier) converts this highly asymmetrical pulse to a nearly Gaussian-shaped output pulse having a width on the order of 1 µs. The shaper also further amplifies the pulse; the peak voltage of the output pulse from the shaper is proportional to the input charge (calibration of the proportionality constant will be discussed below.) If a fast ADC is available then the ADC can directly sample the pulse from the shaper; otherwise the output of the shaper goes to a stretching amplifier, which stretches the Gaussian pulse into a rectangular-shaped output pulse, having a width on the order of 10 µs, which can then be easily sampled by an ADC.

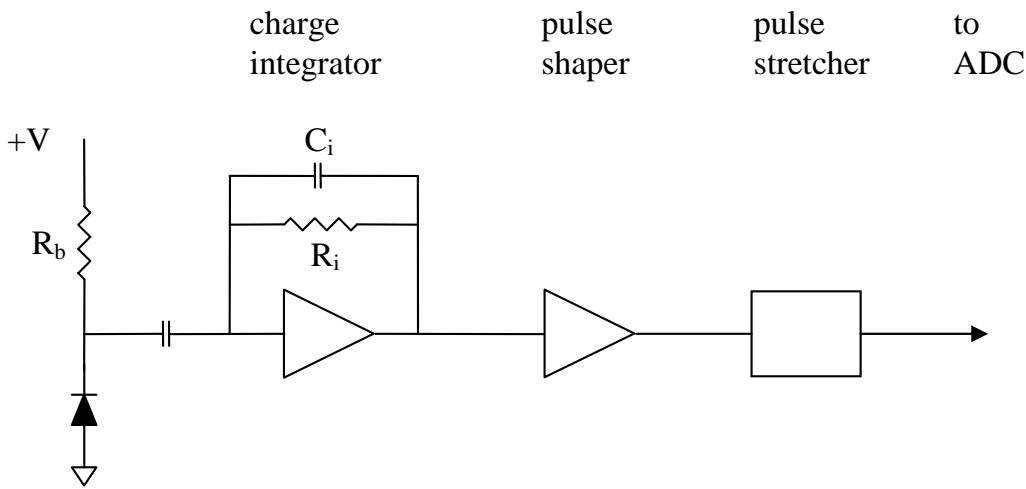

Fig. 15 Circuit diagram for a DC detection system.

Charge-sensitive preamplifiers, spectroscopy amplifiers and pulse stretchers are commercially available from vendors that manufacture electronics for nuclear spectroscopy. Vendors we have used include:

- Amptek Inc; Bedford, MA; www.amptek.com.
- eV Products Inc.; Saxonburg, PA; www.evproducts.com.
- Ortec; Oak Ridge, TN; www.ortec-online.com.
- Canberra Industries; Meriden, CT; www.canberra.com.

A primary objective in building a system is to achieve low electronic noise – a target of 10 dB below the shot noise is suitable. One source of noise is Johnson noise in the resistors. Another is series resistor noise in the field-effect transistor (FET) channel of the charge-sensitive preamp. This noise depends on properties of the FET (such as the carrier transit time through the channel), the input capacitance (usually dominated by the intrinsic capacitance of the detector), and the shaping time of the amplifier. For detailed discussions see [139, 140]. For a fixed FET and shaping time, lowering the input capacitance lowers the noise. For a fixed capacitance, it is possible to choose the proper FET and shaping time to lower the noise. Consulting the specifications of the charge-sensitive preamp can make these choices. Depending on the detector capacitance (typically 10 pF), amplifier gain (typically $10^{-6}$ V/e$^-$) and shaping time (typically 1μs), typical noise levels range from 200–900 electrons rms (root-mean-square). To achieve such performance for single-pulse measurements, the pulse repetition rate must be less than the inverse of the shaping time, preventing operation faster than about 1 MHz with current technology.

The system shown in Fig. 15 contains a single photodiode. Since two detectors are needed for balanced detection, an identical second system is needed. When using two separate detectors once can measure the number of photoelectrons in each detector, $N_1$ and $N_2$, and then use these to calculate the photoelectron sum $N_+ = N_1 + N_2$ and difference $N_- = N_1 - N_2$. As discussed in Sec. 2 above, the scaled difference number yields the quadrature amplitude, which is the quantity of interest in performing OHT. The scaling factor is proportional to the average of the total number of photoelectrons, which is simply given by $\langle N_+ \rangle$. Thus, using two separate detectors makes determination of the scaling factor quite easy; it is determined *in situ* as the quadrature amplitude data is collected.

The price paid for this convenience is that the ADC must be capable of simultaneously recording the outputs from two detectors. It must also be capable of resolving the number of photoelectrons on a given channel, e.g. $N_1$, with a resolution better than the standard deviation of the fluctuations in that channel, (i.e., to better than the square-root of the SNL $\langle (\Delta N_1)^2 \rangle^{1/2} = \langle N_1 \rangle^{1/2}$, where $\Delta N_1 \equiv N_1 - \langle N_1 \rangle$.) For example, suppose a detector has an electronic noise level of 300 electrons rms. In order that the signal dominate this noise, we must have the shot noise fluctuations of the signal much larger than this; $\langle N_1 \rangle^{1/2} = 1200$ photoelectrons rms is sufficient, as this corresponds to a shot-noise variance 12 dB above the electronic noise variance. The average number of photoelectrons needed to achieve this is $\langle N_1 \rangle = 1.44 \times 10^6$. Thus, the resolution of the ADC must be greater than $\langle N_1 \rangle / \langle N_1 \rangle^{1/2} = \langle N_1 \rangle^{1/2}$. In this example a 12-bit digitizer having a resolution of 1 part in 4,096 is not adequate. Typically 16-bit digitizers are required when independent measurements of two detectors are performed.

An alternative to the simultaneous sampling of two channels is first to subtract the photodiode currents, and then amplify and digitize. In this case the single reverse-biased photodiode shown in Fig. 16 is replaced with two photodiodes as shown in Fig. 16. The difference is that instead of detecting two large numbers and then subtracting, the difference number $N_-$ is directly detected, so lower-resolution digitizers are sufficient (12-bit resolution is more than adequate.)

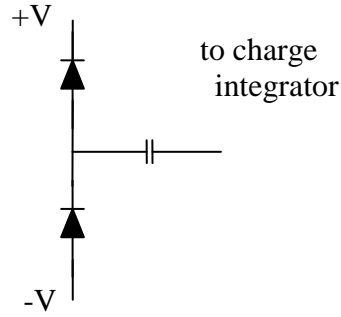

Fig 16 Photodiode electrical configuration for subtraction before integration.

The disadvantage of subtracting first is that it is no longer possible to calibrate the SNL *in situ*. It is necessary first to block the signal beam and measure the variance of the difference number $\langle(\Delta N_-)^2\rangle$ with only the LO incident on the photodetectors. If the (classical) noise fluctuations of the LO are not too large, and the precision of subtraction is adequate to suppress them, then $\langle(\Delta N_-)^2\rangle$ should be equal to the SNL. To verify that this is the case, it is standard practice to block one of the detectors and measure the average number of photoelectrons incident on the other detector $\langle N_1\rangle$. If the detectors are well balanced, the average number of photons hitting the second detector should be the same, so the total number of detected photoelectrons is $2\langle N_1\rangle$. As long as $\langle(\Delta N_-)^2\rangle = 2\langle N_1\rangle$ one is confident that the detectors are operating at the shot-noise limit.

In the above we have been discussing measurements of photoelectron number $N$, however the quantity measured directly by the ADC is a voltage $V$. In order to obtain $N$ from $V$ it is necessary to measure the gain $g$ of the detection system, such that

$$N = gV \qquad . \qquad (4.1)$$

There are two methods to determine the gain, and it is standard practice to use both methods and to ensure that they are consistent.

In the first method the input to the charge-sensitive preamp is replaced by a test capacitor connected to a voltage pulser. By applying a voltage pulse of known height $V_p$, a known charge $Q = CV_p$ is delivered to the amplifier and its output voltage $V$ is measured. The gain of the amplifier is then given simply by

$$g = \frac{Q}{V} = \frac{CV_p}{V} \qquad . \qquad (4.2)$$

The accuracy of this calibration is limited to about 5 or 10% by the accuracy of the calibration of the capacitance.

The second method simultaneously provides measures of the gain and the electronic noise, as well as verifying that the detector is operating at the SNL. If a balanced homodyne detector makes measurements at the SNL, then

$$\langle(\Delta N_-)^2\rangle = \langle N_+\rangle + \sigma_e^2 \qquad , \qquad (4.3)$$

where we have added the variance of the electronic noise $\sigma_e^2$.

The ADC measures the voltage outputs from the two detectors, $V_1$ and $V_2$. We define the sum and difference voltages as $V_+ = V_1 + V_2$ and $V_- = V_1 - V_2$. Using the fact that the measured voltage $V$ is proportional to the photoelectron number, Eq. (4.1), we can rewrite Eq. (4.3) as

$$g^2 \langle (\Delta V_-)^2 \rangle = g \langle V_+ \rangle + \sigma_e^2 \quad , \tag{4.4}$$

or

$$\langle (\Delta V_-)^2 \rangle = \frac{1}{g} \langle V_+ \rangle + \frac{1}{g^2} \sigma_e^2 \quad . \tag{4.5}$$

Thus, by varying the LO pulse energy and plotting $\langle (\Delta V_-)^2 \rangle$ vs. $\langle V_+ \rangle$ we should obtain a linear plot. The slope of the line determines the gain, and the intercept determines the electronic noise. If the plot is not linear then the detector is not operating at the SNL. Data illustrating this calibration procedure can be found in Ref. [62].

### 4.b RF Detection

The majority of balanced detection circuits fall into the class of radio frequency RF detectors. This class was developed for the detection of squeezed light, and have been widely used ever since [28, 58]. Here the current fluctuations from the photodetectors are electrically filtered to within some bandpass about a nonzero center frequency. The filtering is frequently done using either an RF spectrum analyzer or an RF mixer in combination with a filter, and the center frequency is typically in the range of 1 to 10's of MHz. For discussions of some of the technical aspects of operating an RF detector, the reader is referred to articles by Machida and Yamamoto [30], and Wu et al. [141].

One difference between DC and RF detection is in the quantity that is measured. As discussed above, the quantity measured, after scaling, in DC balanced detection is the rotated quadrature amplitude of a particular mode of the signal field

$$\hat{q}_\theta = \frac{1}{\sqrt{2}} \left( \hat{a} e^{-i\theta} + \hat{a}^\dagger e^{i\theta} \right) \quad . \tag{4.6}$$

In RF detection the LO has an optical frequency $\omega$, and the detector current is analyzed with a bandpass filter centered at the radio frequency $\Omega$. The detected RF can come from beating between the LO and signal fields at optical frequencies of $\omega \pm \Omega$. Thus, the detected signal arises from combinations of optical modes at two different frequencies. The measured quadrature operator is then [42]

$$\hat{q}_\theta = \left( \hat{a}_{\omega+\Omega} e^{-i\theta} + \hat{a}^\dagger_{\omega-\Omega} e^{i\theta} \right) + \left( \hat{a}^\dagger_{\omega+\Omega} e^{i\theta} + \hat{a}_{\omega-\Omega} e^{-i\theta} \right) \quad . \tag{4.7}$$

If the LO is pulsed, and therefore not monochromatic, this formula must be generalized by integrating over $\omega$ [67].

### 4.c Other experimental issues

One of the more difficult, but important, aspects of performing OHT is achieving balanced detection at the SNL. If the detector noise is not shot-noise limited then one cannot measure the true quantum state. An excellent test for this is to block the signal beam completely, thus creating a signal in a vacuum state; then experimentally reconstruct this vacuum state and see how well it corresponds to a true vacuum state.

Fluctuations of the LO intensity (beyond the expected Poisson fluctuations) cause the detector noise to rise above the SNL. Balanced detection can largely eliminate the deleterious effects of these fluctuations if everything is working properly. Noise levels above the SNL are often due to poor subtraction between the two detectors. This can be caused by poor alignment of the beams on the detector faces, or by unequal splitting of the LO on the beam splitter. One must ensure that all of the light leaving the beam splitter is collected by the photodiodes. Nearly perfect 50/50 splitting of the LO beam is also necessary to achieve high subtraction efficiency. One method for achieving this is to use a combination of a λ/2-plate and a polarizing beam splitter as a 50/50 beam splitter [142]. Rotation of the λ/2-plate allows one to adjust the splitting ratio of the beam splitter. This issue is more critical for DC detection than for RF detection.

As discussed in Sec. 2, detection quantum efficiency (QE) plays an important role in OHT, as losses degrade information about the measured state. One wants to use the highest-QE detectors possible; photodiodes with efficiencies of over 85% are easily available in the near-IR portion of the spectrum. Furthermore, the homodyne efficiency of the system must be large. Homodyne efficiency is a measure of the overlap of the spatial-temporal field modes of the signal and LO beams [see Eq. (3.6)]. This can be maximized by careful alignment, however, for signal fields created by nonlinear processes it is nearly impossible to perfectly overlap the signal and LO modes [143-145]. One method of getting around this is to generate an LO using the same nonlinear process, so that it is matched to the signal mode [146]. Array detection can also be used to circumvent some of the losses due to mode-mismatch; we discuss this alternative in the following section.

When studying light with a nonzero mean amplitude (such as coherent states or bright squeezed states) using the DC whole-pulse technique, it is crucial to achieve extremely precise balancing between the two detection channels. (This is less important when using RF techniques.) In order to do this, both stages of the two detection chains needs to be equalized–the effective quantum efficiency (QE) of the two photodiodes, and the overall electronic gain must be the same for each channel. The effective QE's can be matched by introducing a small variable optical loss in front of the detector with the highest intrinsic QE (e.g., by using a rotated glass Brewster plate).

A procedure for setting the gain of the amplifiers used with the detectors is as follows [147]. As describe above, voltage pulses of known amplitude are used in conjunction with capacitors to introduce a known charge into the input of the amplifiers. Since two detectors are being calibrated simultaneously, an input pulse is sent through a 50 Ω voltage divider with a nominal splitting of 50% to produce two outputs. The voltage divider can be adjusted to change the ratio of the outputs from 1:1 to slightly more or less. High quality ceramic chip microwave grade capacitors are used to obtain proper operation over large bandwidths. Semi-rigid SMA cables and connectors are used to provide stability of the intrinsic capacitance of the connectors themselves when the connectors are subjected to mechanical strains such as when the cables are connected and disconnected. The intrinsic capacitance of the SMA cables and connectors changes much less than the capacitance of the test capacitors. Since the gain of the preamps change with the capacitance at the input, the photodiodes are left in place during the electronic calibration.

If we have two charge pulses, one from each test capacitor, with charges $Q_1$ and $Q_2$, the voltage measured on each channel at the computer will be proportional to the charge contained in the pulse. Therefore, $V_1 = \alpha Q_1$, $V_2 = \beta Q_2$, where $V_1(V_2)$ is the voltage measured by the computer on channel 1(2) with $\alpha(\beta)$ the overall conversion gain from the charge to the voltage for channel 1(2). Note that $Q_1$ and $Q_2$ are derived from the same voltage pulse. If we interchange the inputs to the charge preamplifiers, then the voltages at the computer will be given by $V_1 = \alpha Q_2$, $V_2 = \beta Q_1$. If in both cases the difference between the voltages is equal to zero, i.e. $V_1 - V_2 = \alpha Q_1 - \beta Q_2 = \alpha Q_2 - \beta Q_1 = 0$, then it is necessary that $\alpha = \beta$ and $Q_1 = Q_2$.

Experimentally, to set the conditions such that $V_1 - V_2 = \alpha Q_1 - \beta Q_2 = \alpha Q_2 - \beta Q_1 = 0$ are satisfied, the pulse generator is set to a voltage that provides approximately $10^6$ electrons to each preamplifier, and the gains are adjusted so that the difference between the two channels at the computer is zero. At this point, we do not know if $\alpha = \beta$, because we do not know if the input charges are the same.

Next the inputs to the preamplifiers are switched so that the difference measured at the computer is $V_1 - V_2 = \alpha Q_2 - \beta Q_1$. Assuming that for the previous step, $Q_1 \neq Q_2$, we then have that $V_1 - V_2 \neq 0$. The 50 Ω variable voltage splitter is now adjusted, changing the amount of charge sent to each preamplifier until the difference measurement is zero, $V_1 - V_2 = 0$. The inputs to the preamplifiers are switched back to the original configuration and the voltage difference is once again measured. The process of adjusting the gain, switching the inputs, adjusting the voltage splitter, and switching the inputs back is iterated until the difference measurement is as close to zero as possible for the configuration of both inputs, with no adjustments to the gains or the voltage splitter.

We can calculate how closely the gains can be set equal to each other given a finite (but small) difference number instead of a difference number of zero as assumed above. The above relations are replaced by $V_1 - V_2 = \alpha Q_1 - \beta Q_2 = n_{diff\,1}$ and $V_1 - V_2 = \alpha Q_2 - \beta Q_1 = n_{diff\,2}$ for the two connection configurations. It is straightforward to show that for small difference numbers, α and β can be set equal to within a precision $(n_{diff\,1} + n_{diff\,2})/n_{tot}$ where $n_{tot}$ is the total number of photoelectrons. Experimentally, for a total number of electrons $n_{tot} = 10^6$ it is relatively easy to achieve difference numbers $n_{diff\,1}, n_{diff\,2} \approx 10^2$. Thus it is possible to achieve $\alpha = \beta$ to within 1 part in $10^4$.

## 5. Array Detection

The use of array detectors was suggested by Raymer et al. for measuring quantum phase distributions [148]. Their use for measuring quantum states was analyzed by Beck [149] and by Iaconis et al. [150]. Experiments to measure density matrices [31, 32] and Husimi distributions (Q-functions) [33] using arrays have been performed. We note that the use of single-photon counting arrays has been suggested for state measurement [151], but this falls under the category of photon-counting methods [152, 153], not OHT; the arrays required for this are quite different from the arrays used in experiments to date.

Array detection is a form of DC detection technology in which the individual detectors used in the balanced homodyne detector are replaced by array detectors. These arrays have many pixels, and resolve the transverse intensity profiles of the beams that illuminate them. The ability to resolve the transverse structure offers several advantages: increased effective detection efficiency, the ability to measure many spatial modes simultaneously, the ability to find the mode which meets a particular definition of "optimal", and even the ability to perform unbalanced homodyne measurements at the SNL (i.e., to use a single output port from a beam splitter instead of subtracting the outputs of two ports.) A disadvantage is that array detection is slow using current technology.

### 5.a Array Detection of Spatial Modes

In Sec. 2 we considered an expansion of the signal and local oscillator fields in terms of spatial-temporal modes $v_k(\underline{x},0,t)$. Here we are mainly interested in the spatial part of the mode function, and for simplicity will assume that the spatial and temporal parts of the mode function factorize. We'll furthermore assume that the signal and LO temporal mode functions, as well as their polarizations, are perfectly matched. In this case we can perform the time integral in Eq. (a.4), and with proper normalization we are left with

$$\hat{N}_- = -i \int_{Det} d^2x \; v^*_L(\underline{x},0) \; \hat{\Phi}_S^{(+)}(\underline{x},0) \; + \; h.c. \quad . \tag{5.1}$$

After time integration the orthogonality condition for the mode functions [Eq.(2.11)] becomes

$$\int_{Det} d^2x \; v^*_k(\underline{x},0) \; v_m(\underline{x},0) = \delta_{k\,m} \quad . \tag{5.2}$$

In Sec. 2 we expanded the LO and signal modes in the same set of mode functions, in order to demonstrate that the signal is projected onto the mode of the LO when using single detectors. This is not the case with array detection, however, so it is convenient to expand the LO and signal using separate mode functions. For the LO modes we'll use the mode functions $v_k(\underline{x},0)$ described above, while for the signal we'll use $w_k(\underline{x},0)$. The $w_k$'s satisfy the same orthogonality condition as the $v_k$'s:

$$\int_{Det} d^2x \; w^*_k(\underline{x},0) \; w_m(\underline{x},0) = \delta_{k\,m} \quad . \tag{5.3}$$

While the $v_k$'s are orthogonal and the $w_k$'s are orthogonal, the $v_k$'s are not in general orthogonal to the $w_k$'s. In the new basis, the mode expansion of the photon flux of the signal field [Eq. (2.9)] becomes

$$\hat{\Phi}_S^{(+)}(\underline{x},0) = i \sum_k \hat{a}_k w_k(\underline{x},0) \qquad . \qquad (5.4)$$

In balanced array detection the individual detectors shown in Fig. 1 are replaced by arrays. The arrays are made up of individual detectors (pixels) that spatially resolve the intensity of the light that illuminates the array. Note that the area of integration in Eqs. (5.1)-(5.3) is over the entire area of the array, $A_a$. Thus, the difference number in Eq. (5.1) refers to the difference between the total number of photons striking array 1 and the total number striking array 2.

We now need to consider what is measured at each individual pixel of the array. The difference number $\hat{N}_{-j}$ is obtained by subtracting the output from pixel $j$ of array number 2 from the corresponding pixel on array 1. This difference number is found by replacing the integration over the entire array, Eq. (5.1), by integration only over the area of pixel $j$

$$\hat{N}_{-j} = -i \int_{pixel\ j} d^2x\ v^*_L(\underline{x},0)\ \hat{\Phi}_S^{(+)}(\underline{x},0)\ +\ h.c. \qquad . \qquad (5.5)$$

Substituting Eq. (5.4) into Eq. (5.5) we obtain

$$\hat{N}_{-j} = \sum_k \hat{a}_k \hat{c}_L^\dagger \int_{pixel\ j} d^2x\ v^*_L(\underline{x},0)\ w_k(\underline{x},0)\ +\ h.c. \qquad . \qquad (5.6)$$

We now assume that the LO is in a plane-wave coherent state. The properly normalized plane-wave mode is

$$v_m(\underline{x},0) = (A_a)^{-1/2} \qquad . \qquad (5.7)$$

As in Sec. 2, the fact that the LO is a large-amplitude coherent state means that we can replace the amplitude $\hat{c}_L$ by $|\alpha_L|e^{i\theta}$. For this approximation to be valid $|\alpha_L|$ must be sufficiently large that each pixel in the array is illuminated by a large-amplitude coherent state. With this assumption, Eq. (5.6) becomes

$$\hat{N}_{-j}(\theta) = \frac{|\alpha_L|}{(A_a)^{1/2}} \sum_k \int_{pixel\ j} d^2x\ \hat{a}_k w_k(\underline{x},0) e^{-i\theta}\ +\ h.c. \qquad . \qquad (5.8)$$

If the spatial variations of the signal field are well resolved by the array, then the relevant mode amplitudes are approximately constant over the dimensions of a pixel. We can then integrate over the pixel area and obtain

$$\hat{N}_{-j}(\theta) = \frac{|\alpha_L| A_p}{(A_a)^{1/2}} \sum_k \left( \hat{a}_k w_k(\underline{x}_j,0) e^{-i\theta} + \hat{a}_k^\dagger w^*_k(\underline{x}_j,0) e^{i\theta} \right) \qquad , \qquad (5.9)$$

where $A_p$ is the area of an individual pixel and $\underline{x}_j$ is the location of pixel $j$.

Since discreet pixels are being used, it is convenient to express the normalization condition Eq. (5.3) in terms of a discreet sum as

$$A_p \sum_j w^*_k(\underline{x}_j,0) w_m(\underline{x}_j,0) \cong \delta_{km} \qquad . \qquad (5.10)$$

Furthermore, if at least one of the mode functions [e.g. $w_m(\underline{x}_j,0)$] is real, then taking the complex conjugate of Eq. (5.10) yields

$$A_p \sum_j w_k(\underline{x}_j,0) w_m(\underline{x}_j,0) \cong \delta_{km} \quad (w_m \text{ real}) \tag{5.11}$$

Multiplying both sides of Eq. (5.9) by $w_m(\underline{x}_j,0)$, summing over *j*, and using Eqs. (5.10) & (5.11) yields

$$\sum_j \hat{N}_{-j}(\theta) w_m(\underline{x}_j,0) = \frac{|\alpha_L|}{(A_a)^{1/2}} \sum_k \left( \hat{a}_k e^{-i\theta} \delta_{km} + \hat{a}_k^\dagger e^{i\theta} \delta_{km} \right) \tag{5.12}$$

Performing the sum over *k*, and rearranging demonstrates that the quadrature amplitude for the signal in mode *m* is given by [149]

$$\hat{q}_{m\theta} = \frac{1}{\sqrt{2}} \left( \hat{a}_m e^{-i\theta} + \hat{a}_m^\dagger e^{i\theta} \right)$$
$$= \frac{1}{|\alpha_L|} \left( \frac{A_a}{2} \right)^{1/2} \sum_j \hat{N}_{-j}(\theta) w_m(\underline{x}_j,0) \tag{5.13}$$

The detector itself yields measurements of $\hat{N}_{-j}(\theta)$, while according to Eq. (5.13) the quadrature amplitude $\hat{q}_{m\theta}$ corresponding to the measured mode *m* is determined after all the data has been collected by summing the measured values of $\hat{N}_{-j}(\theta)$ with a weighting factor given by the mode function $w_m(\underline{x}_j,0)$. Since an array detector is capable of making measurements of the quadrature amplitude, these measurements can be used to reconstruct the quantum state of the signal in mode *m*, as described above in Sec. 3.

The fact that an array detector can measure the state of an optical field is not surprising. What probably is surprising is that in this detection scheme the mode functions of the measured mode $w_m(\underline{x}_j,0)$ and the plane-wave LO mode are not the same, but this mode-mismatch does *not* reduce the effective detection efficiency of the measurements as it would when using standard point-like detectors. Any properly normalized, real mode function $w_m(\underline{x}_j,0)$ can be used in Eq. (5.13), and the measured quadrature amplitude is not reduced by a factor proportional to the overlap of the signal and LO modes [as in Eq. (3.6).] . In Ref. [31] array detection was found to be over 40 times more efficient than standard detection for a particularly poorly matched set of LO and signal modes. One limitation is that the measured mode $w_m(\underline{x}_j,0)$ must be real, that is, must have a constant phase across its profile. This ensures that the quadrature amplitude in Eq. (5.13) is real.

The mode function enters into the determination of the quadrature amplitudes during the data analysis, after all the data has been collected. Thus, by choosing different mode functions it is possible to determine the quadrature amplitudes of many different spatial modes for any given set of measurements $\hat{N}_{-j}(\theta)$ [31]. Despite the fact that the states of many modes may be measured simultaneously, it is not possible to use this technique directly to measure the *joint* quantum state of these modes. This is because all of the modes are measured with the same rotation angle (phase shift) θ; to determine the joint quantum state each mode must have its own independently adjustable phase [121, 154].

In Fig. 17 we plot data from Ref. [31] showing the corrected difference number $\hat{N}_{-j}(\theta) - \left\langle \hat{N}_{-j}(\theta) \right\rangle_{vac}$ as a function of pixel number observed across a one-dimensional array detector; details about subtraction of the vacuum average are deferred until Sec. 5.d. In Fig. 17 the signal field is in a weak coherent state having a mean of approximately 1 photon; Fig. 17(a) shows data collected on a single exposure, while Fig. 17(b) shows data averaged over 200 exposures. The two curves in each figure differ in that each curve corresponds to a different value of the LO phase; the phase difference between them is π. For this experiment the weak signal field occupied a field mode in which amplitude varied linearly with position, and the data in Fig. 17 confirm this. Notice that there is a π phase shift in the middle of the beam (the difference counts tend to be negative for half the beam, and positive for the other half). Changing the LO phase by π causes the slope of the curves in Fig. 17 to invert (positive difference counts become negative and vice versa), as expected.

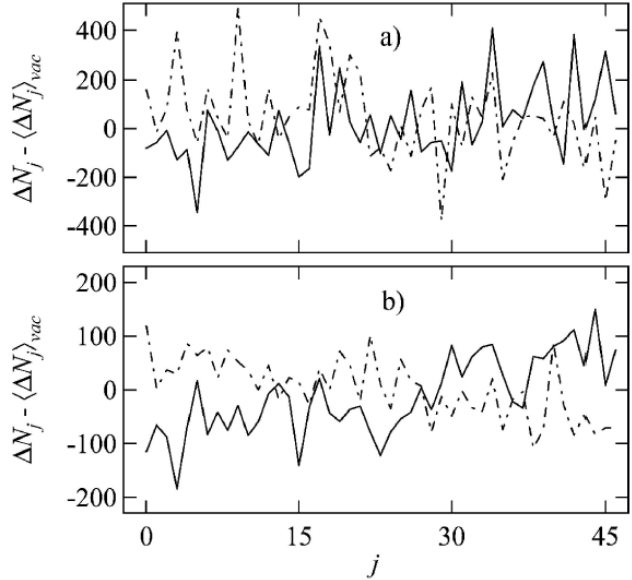

Fig. 17 The corrected difference number is plotted as a function of pixel number for a signal mode in a coherent state with a mean of approximately 1 photon: a) shows data for a single exposure, while b) shows an average of 200 exposures. The two curves in each figure correspond to two values of the local oscillator phase that differ by π. From Ref. [31].

Figure 17 is a dramatic illustration of interference at the single-photon level. While these curves contain large noise (due to the shot noise of the LO and imperfect subtraction of the vacuum difference level), they can clearly be seen to have opposite slopes. An average of one photon in the signal beam, even on single shots as shown in Fig. 17(a), can lead to macroscopic differences in the detected signal across *many* pixels of the array. The single signal photon acts as a "traffic cop," determining where the millions of photons in the LO beam strike the array.

**5.b Optimization of the Measured Mode**

Since it is possible to measure many different modes for a given set of data, it is natural to consider whether there is a procedure for finding the mode that optimizes the measurement of a particular quantity. Dawes et al. have shown that it is indeed possible to optimize the measurement of quantities that are quadratic in the field operators [32].

For example, consider the case of trying to find the mode that maximizes the average number of detected signal photons. In terms of the quadrature amplitudes, the average photon number of mode *m* is given by

$$\langle \hat{N}_m \rangle = \frac{1}{2\pi} \int_0^{2\pi} d\theta \langle \hat{q}_{m\theta}^2 \rangle - \frac{1}{2} \qquad . \qquad (5.14)$$

Substituting Eq. (5.13) into Eq. (5.14) leads to

$$\left\langle \hat{N}_m \right\rangle = \frac{A_a}{2|\alpha_L|^2} \mathbf{w}^T \cdot \underline{\mathbf{M}} \cdot \mathbf{w} - \frac{1}{2} \quad , \tag{5.15}$$

where we have introduced vector notation. Here $\underline{\mathbf{M}}$ is the correlation matrix for the difference photocounts, averaged over the phase of the local oscillator:

$$\underline{\mathbf{M}}_{jj'} = \frac{1}{2\pi} \int_0^{2\pi} d\theta \left\langle \hat{N}_{-j}(\theta) \hat{N}_{-j'}(\theta) \right\rangle \quad , \tag{5.16}$$

and $\underline{\mathbf{w}}_j = w_m(\underline{x}_j, 0)$ is a vector composed from the values of the mode function taken at the pixels of the array detector.

Our goal is to find the vector $\mathbf{w}$ that maximizes $\left\langle \hat{N}_m \right\rangle$. As the second term on the right-hand side of Eq. (5.14) is constant, this task is in turn equivalent to finding the eigenvector of $\underline{\mathbf{M}}$ corresponding to its maximum eigenvalue. Once the optimal mode has been determined it can then be substituted for $w_m(\underline{x}_j, 0)$ in Eq. (5.13), and the quadrature amplitudes corresponding to this mode can be computed. These amplitudes can then be used to determine the quantum state of the field. This technique was experimentally demonstrated by Dawes et al. in Ref. [32].

We note that we are able to use standard numerical methods for solving symmetric eigenvalue problems because the quantity used as the optimization criterion is quadratic in the quadrature amplitude operators. Similar methods can be used to optimize other quadratic operators (e.g., it is possible to find the mode that maximizes the amount of detected squeezing.) In a general case the optimization criterion can be a highly nonlinear function, which makes the optimization problem significantly more complicated.

### 5.c Joint Q-Function of Many Modes

In the above-described experiments with arrays it was possible to obtain the full quantum state of many modes simultaneously, but it was not possible to obtain the joint state (i.e., the correlations between modes were lost.) However, it is possible to use array detection to obtain information about correlations between the modes. Indeed, the experimental procedure for extracting this information is less complicated than the previously described array experiments, as the use of balanced detection is not necessary (only a single array is needed.) The price paid for joint information is that one does not measure the Wigner function or the density matrix of the field modes, but instead the joint Q-function of the modes.

The Q-function, or Husimi function, is a quantum mechanical, phase-space, quasi-probability distribution; it is positive definite and may be used to calculate quantum expectation values of antinormally ordered operators [1, 155]. It is equal to the state's Wigner function convolved with the Wigner function of the vacuum state. In principle the Q-function contains all information about the quantum mechanical state of a system. However, to extract the density matrix from the Q-function it is necessary to perform a numerical deconvolution, which is impractical with real experimental data. Despite this limitation, it is possible to calculate low-order moments of anti-normally ordered operators using the Q-function (e.g., to obtain moments of photon number and a suitably defined phase.)

Measurement of the joint Q-function of many temporal modes was demonstrated in Ref. [33]. In this experiment it was possible to obtain shot-noise limited operation without balancing because the signal and LO occupied temporally distinct modes, so that classical noise on the LO could be separated out. Needed interference between the LO and signal was obtained by making measurements in the frequency domain where they overlapped, and the noise was eliminated during data processing by Fourier transforming back into the temporal domain.

Following the analysis in Refs. [33] & [150], we consider a time window consisting of $2M+1$ temporal modes $\hat{b}_k$. The signal and LO fields are separated by a time delay. For the purposes of this analysis we will thus assume that the LO occupies the $2J+1$ temporal modes near the center of our time window ($J < M/2$). The signal occupies the temporal modes after the LO, and the modes before the LO are empty. In order to make this distinction between these modes more clear, we rewrite the mode operators as

$$\hat{b}_k = \begin{cases} \hat{b}_k^{(vac)} & -M \leq k < -J \\ \hat{b}_k^{(LO)} & -J \leq k \leq J \\ \hat{b}_k^{(S)} & J < k \leq M \end{cases}, \qquad (5.17)$$

where the superscripts refer to vacuum, LO, or signal mode operators.

In the experiment the LO and signal pulses are measured by an array detector at the back focal plane of a grating spectrometer. The number operator for the measured spectral modes is $\hat{N}_j = \hat{a}_j^\dagger \hat{a}_j$, where annihilation operators $\hat{a}_j$ can be expressed as the Fourier transform of the temporal mode operators as

$$\hat{a}_j = \frac{1}{\sqrt{(2M+1)}} \sum_k \exp[i 2\pi j k / (2M+1)] \hat{b}_k \qquad . \qquad (5.18)$$

The quantity of primary interest in the experiments corresponds to the Fourier transform of $\hat{N}_j$,

$$\hat{K}_l \equiv \sum_j \exp[-i 2\pi l j / (2M+1)] \hat{N}_j \qquad . \qquad (5.19)$$

Equations (5.17)-(5.19) can be combined to express $\hat{K}_l$ in terms of temporal mode operators $\hat{b}_k$. Terms of second order in operators corresponding to the weak fields $\hat{b}_k^{(S)}$ and $\hat{b}_k^{(vac)}$ are discarded. Furthermore, since the modes of the LO pulse are in large-amplitude coherent states, the dominant contributions are retained if we replace the LO mode operators $\hat{b}_k^{(LO)}$ by their corresponding coherent-state amplitudes $\beta_k$. The terms that contribute to the summations in the expression for $\hat{K}_l$ depend on the value of $l$; for $l > 2J$ we find

$$\hat{K}_l = \sum_{k=-J}^{J} \left( \beta_k^* \hat{b}_{k+l}^{(S)} + \beta_k \hat{b}_{k-l}^{\dagger(vac)} \right) \qquad . \qquad (5.20)$$

If the LO occupies only a single ($k = 0$) temporal mode, then Eq. (5.20) simplifies to

$$\hat{K}_l = \beta_0^* \hat{b}_l^{(S)} + \beta_0 \hat{b}_{-l}^{\dagger(vac)} \qquad . \qquad (5.21)$$

Notice that terms quadratic in the LO amplitude (i.e., the terms normally eliminated by subtraction when using balanced homodyne detection) are absent from Eqs. (5.20) & (5.21). This is because these terms are located near $l=0$, not near the terms of interest, $l>2J$. Thus, LO noise is effectively removed, without the need to perform balanced detection.

Measurement of $\hat{K}_l$ returns a complex number, which can be interpreted as a measurement of the signal mode plus an added vacuum noise contribution [Eq. (5.21)]. Real and imaginary parts of the measurement correspond to simultaneous measurement of the quadrature amplitudes $q_l$ and $p_l$. The price paid for simultaneous measurement of noncommuting observables is the presence of the additional vacuum noise, as was first pointed out by Arthurs and Kelly [156,1]. A similar situation arises in heterodyne detection [103].

By histogramming the measured values of $q_l$ and $p_l$ one creates a joint probability distribution, which in the limit of a large number of samples tends to the Q-distribution for the field quadratures $Q(q_l, p_l)$. Since the quadrature amplitudes for all values of $l$ are measured simultaneously, joint Q-distributions for multiple modes can be created. Figure 18(a), taken from Ref. [33], shows the measured Q-function of a single temporal mode of an optical pulse. This is a two-dimensional histogram of measured quadrature amplitudes on 14,000 shots. The signal and LO beams came from the two arms of a Michelson interferometer; the signal arm also contained 1.5 cm of glass that added dispersion to the signal pulse; the signal pulse was thus stretched and chirped. The path-length difference between the two arms was not stabilized; this randomized the signal phase producing non-Gaussian Q-functions. The Q-function of Fig. 18(a) is largely circular, however there is still a peak in the distribution indicating that the phase was not completely randomized.

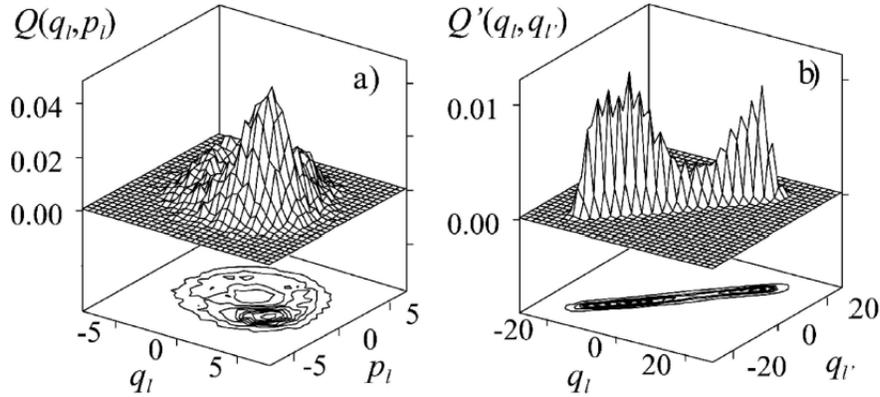

Fig 18 The measured Q-functions of a chirped signal pulse with a random phase: a) shows the Q-function of a single mode, while b) shows correlations between 2 modes. [33]

Two-mode distributions have the form $Q(q_l, p_l, q_{l'}, p_{l'})$; these four-dimensional distributions are difficult to display graphically, so correlations between modes are usually displayed in terms of the joint distribution of the $q$ quadratures

$$Q'(q_l, q_{l'}) = \iint Q(q_l, p_l, q_{l'}, p_{l'}) dp_l dp_{l'} \quad , \qquad (5.22)$$

Figure 18(b) shows correlations between the $q$-quadratures of two modes in terms of joint distributions $Q'(q_l, q_{l'})$. This figure shows correlations between two temporal modes that were both near the peak of the pulse, and consequently had nearly the same amplitude and phase. This joint distribution lies mainly along a line whose slope is 1, indicating strong positive correlations between the $q$-quadratures. This is what one would expect for two modes whose relative phases are nearly the same, but whose absolute phases are random.

In the experiment of Ref. [33], the exposure time of the array was 300 ms, so each "shot" was actually composed of millions of pulses, since the laser had a repetition rate of 82 MHz. Any noise at frequencies of 1/(300 ms) and higher was then integrated and averaged over. The CCD array used had a very slow read time, so the repetition rate of the measurements was approximately 1/2 Hz, yielding an experimental run of nearly eight hours. Noise due to slow drift of the interferometer phase over this time was the dominant contribution to the shapes of the measured distributions in Fig. 18. Furthermore, since many pulses were integrated on each shot, the experiment did not measure the statistics of temporal slices of an ensemble of single pulses. The measured temporal mode was a "super"-mode, representing the corresponding time slices of millions of pulses. In principle it is not necessary to average over millions of pulses; a laser pulse repetition rate slower than the inverse CCD exposure time would lead measurements of an ensemble of single pulses

## 5.d Technical Considerations

As with ordinary single detectors, one wants array detectors to have high quantum efficiency and low electronic noise. Since arrays are made of many adjacent pixels, one factor that contributes to the efficiency is the "fill factor," which is a measure of the fraction of the detector area that is sensitive to light. Fill factors of less than 100% can be due to gaps between individual pixels. For example, interline-transfer charge-coupled device (CCD) detectors have alternating rows of photosensitive pixels and rows of nonphotosensitive areas that are used to read the charge out of the array. Such CCD's have fill factors of only 25—75 %, and hence are not suitable for quantum optics applications. Full-frame CCD's, on the other hand, have fill factors of 100%. The trade off here is that since charge is transferred from one pixel to the next in order to be read out, charge smearing occurs if the array is exposed to light during readout. This means that an external shutter is necessary to block the light during readout.

CCD's come in two basic types: front-illuminated and back-illuminated. The polysilicon gate structure used to read the charge out of the array is placed on the front surface of the array. In front-illuminated devices light is incident through this gate structure. Light is absorbed and reflected by the polysilicon, reducing the quantum efficiency. In back-illuminated CCD's the backside of the silicon wafer is thinned to a thickness of about 20 μm, and light is incident through the back. There is no gate structure in the way of the light, so the quantum efficiency of back-illuminated devices can exceed 90%. However, the 20 μm layer of silicon can act as an etalon, producing unwanted fringes in the acquired image; this effect is especially pronounced in the near IR. The latest generation of back-illuminated CCD's has been engineered to greatly reduce this etaloning effect.

Electronic noise in CCD's comes in the form of dark noise and readout noise. Scientific grade CCD cameras can be cooled to temperatures of -100$^o$C using either thermoelectric or liquid nitrogen cooling, and at this temperature dark noise is essentially non-existent (on the order of 1 e$^-$/pixel/hr). Readout noise can be less than 10 e$^-$/pixel rms, depending on the readout rate—higher readout rates have larger noise.

Most scientific grade CCD's have 16-bit resolution ADC's. These CCD's are frequently used for Raman or fluorescence spectroscopy, and can be purchased from most vendors that sell spectrometers. Some vendors that we are aware of are:

- Roper Scientific Inc., Trenton, NJ; www.roperscientific.com
- Andor Technology, South Windsor, CT; www.andor-tech.com

As with single detectors, it is necessary to ensure that the detection is shot-noise limited. This is done in the same manner as described in Sec. 4.a. The detector is illuminated by an LO beam of varying intensity and the variance of the measured difference number between pairs of pixels is plotted versus the mean. Detection at the SNL yields a linear plot. The slope and intercept determine the gain and electronic noise, and these should agree with the specifications of the manufacturer. This calibration can be done either pixel-by-pixel, or for the sum of the outputs from a large number of pixels.

In balanced array detection it is extremely important to register properly the individual pixels detecting the two beams illuminating the array to obtain high-efficiency subtraction and hence good classical noise reduction. This is one of the more difficult parts of array experiments. If the outputs are not properly registered, operation at the SNL is not obtained. A detailed description of one convenient registration procedure is given in Ref. [31].

Once the pixels have been registered the detector needs to be balanced as well is possible. If the signal field is blocked (i.e., the signal mode entering the detector is in the vacuum state) then the average difference number for each pixel should be zero: $\langle N_{-j}(\theta) \rangle_{vac} = 0$, where the subscript indicates that the signal is in the vacuum state. This is extremely difficult to achieve for every pixel simultaneously. To eliminate the effects of offsets in the measured difference number for individual pixels, it is necessary in practice to subtract these offsets. Thus, in Eq. (5.13) one uses the corrected difference number $N_{-j}(\theta) - \langle N_{-j}(\theta) \rangle_{vac}$ in place of $N_{-j}(\theta)$ when calculating quadrature amplitudes. The measured background and signal levels are obtained in the experiment by alternately blocking and unblocking the signal with a shutter.

In unbalanced array detection there is no need to register or balance pixels, and operation at the SNL is easily obtained, even with LO fluctuations of 15 % peak-to-peak [33].

## 6. Conclusions

Quantum state tomography of optical fields has come of age. Many theoretical algorithms exist for converting measured quadrature amplitudes into information about the quantum state. Numerous experiments have been performed which demonstrate the utility of these algorithms (see Table 1.) These experiments have used several different detection technologies (DC, RF, array), and have measured quantities ranging from Wigner functions to photon number and phase distributions. Techniques have been developed to measure two or more mode systems, allowing for the measurement of temporal or polarization correlations between modes.

Ultrafast linear optical sampling has been a spin off of state measurement technology. The ability of a balanced homodyne detector to perform time-resolved measurements of weak fields is very important from a practical perspective. Work in this area is really just beginning, but we expect it to have a bright future.

**Acknowledgements**

We wish to thank our many collaborators who contributed so much to our work in the area of QST: Matt Anderson, Ethan Blansett, Howard Carmichael, Jinx Cooper, Adel Faradani, Andy Funk, Tamas Kiss, Ulf Leonhardt, Dan McAlister, Mike Munroe, Thomas Richter, Dan Smithey, Ian Walmsley. Andrew Dawes, Konrad Banaszek, and Christophe Dorrer. A. Funk made a significant contribution to the writing of Section 4.c. The work in our groups has been supported by the National Science Foundation and by the Army Research Office.

## Appendix A

Define

$$S_{ml} = c \int_0^T dt \int_{-\infty}^{\infty} d^2x \, \underline{v}^*_m(\underline{x},0,t) \cdot \underline{v}_l(\underline{x},0,t) =$$

$$c \int_0^T dt \int_{-\infty}^{\infty} d^2x \sum_j C^*_{mj} \underline{u}^*_j(\underline{x},0) \exp(i\omega_j t) \cdot \sum_i C_{li} \underline{u}_i(\underline{x},0) \exp(-i\omega_i t) \qquad (A.1)$$

Using plane-wave modes, $\underline{u}_j(\underline{r}) = V^{-1/2} \underline{\varepsilon}^{(j)} \exp(i\underline{k}_j \cdot \underline{r})$, with $V = L_z L_y L_x$, gives

$$S_{ml} = \sum_{j,i} C^*_{mj} C_{li} \, \delta_K(\underline{k}_{Tj} - \underline{k}_{Ti}) \frac{c}{L_z} \int_0^T dt \, \exp[i(\omega_j - \omega_i)t] \qquad (A.2)$$

where $\delta_K(\underline{k}_{Tj} - \underline{k}_{Ti})$ is a Kronecker delta indicating that the transverse components of $\underline{k}_j$ and $\underline{k}_i$ must be equal. In the paraxial approximation ($k_x, k_y \ll k_z$) the frequencies are

$\omega_j/c \cong k_{zj} + (k_{Tj}^2)/2k_{zj}$ and $\omega_i/c \cong k_{zi} + (k_{Ti}^2)/2k_{zi}$, where $k_{zj} = j 2\pi/L_z$ ($j = 1,2,3...$) and $cT = L_z$. This gives (with the constraint $\underline{k}_{Tj} = \underline{k}_{Ti}$),

$$\frac{c}{L_z} \int_0^T dt \, \exp[i(\omega_j - \omega_i)t] \cong \begin{cases} 1 & (j=i) \\ (\theta_0/2)^2 & (j \neq i) \end{cases} \qquad (A.3)$$

where $\theta_0 = k_{Tj}/k_z \ll 1$ is the angle of the modes' propagation vector from the normal to the $z = 0$ plane. For typical laser beam divergences of $\theta_0 \approx 10^{-3}$ this deviation from zero is small, making the time integral (combined with the transverse integral) behave like a Kronecker delta $\delta_{ji}$. This then leads to

$$S_{ml} = \sum_j C^*_{mj} C_{lj} = \delta_{ml} \qquad (A.4)$$